\pdfoutput=1

\documentclass[
 preprint, pre,
 showpacs,
 amsmath,amssymb,
 superscriptaddress,
 aps, 11pt]{revtex4-1}

\usepackage{amsmath}
\usepackage{amssymb}
\usepackage{dcolumn}
\usepackage{bm}
\usepackage{hyperref}
\usepackage{verbatim}
\usepackage[version=4]{mhchem}

\usepackage{wasysym}
\usepackage{verbatim}
\usepackage{graphics}
\usepackage{lipsum}
\usepackage{graphicx}

\usepackage{subcaption}
\usepackage{wrapfig}
\usepackage{epsfig}
\usepackage{float}
\usepackage{array}
\usepackage{psfrag}
\usepackage{color}
\usepackage{bbold}

\definecolor{azure}{rgb}{0.0, 0.5, 1.0}
\definecolor{asparagus}{rgb}{0.53, 0.66, 0.42}
\definecolor{ballblue}{rgb}{0.13, 0.67, 0.8}
\definecolor{sgreen}{rgb}{0.0, 0.8, 0.35}
\definecolor{darkgreen}{rgb}{0.0, 0.5, 0.0}
\definecolor{sred}{rgb}{0.9, 0.6, 0.4}
\usepackage{textcomp}

\usepackage{tikz}

\newcommand{\dd}{\text{d}}

\newcommand{\vect}[1]{\boldsymbol{#1}}

\begin{document}

\title{
Controlling composition of coexisting phases via
molecular transitions
}

\author{Giacomo Bartolucci}
\affiliation{Max Planck Institute for the Physics of Complex Systems,
N\"{o}thnitzer Strasse~38, 01187 Dresden, Germany}
\affiliation{Center for Systems Biology Dresden,  Pfotenhauerstrasse~108, 01307 Dresden, Germany}

\author{Omar Adame-Arana}
\affiliation{Department of Chemical and Biological Physics, Weizmann Institute of Science, Rehovot 76100, Israel}
\author{Xueping Zhao}
\affiliation{Max Planck Institute for the Physics of Complex Systems,
N\"{o}thnitzer Strasse~38, 01187 Dresden, Germany}
\affiliation{Center for Systems Biology Dresden,  Pfotenhauerstrasse~108, 01307 Dresden, Germany}

\author{Christoph A.\ Weber\footnote{\label{CA}Corresponding author.}}
\affiliation{Max Planck Institute for the Physics of Complex Systems,
N\"{o}thnitzer Strasse~38, 01187 Dresden,
Germany}
\affiliation{Center for Systems Biology Dresden,  Pfotenhauerstrasse~108, 01307 Dresden, Germany}

\date{\today}

\begin{abstract}
Phase separation and transitions among different molecular states are ubiquitous in living cells. 
Such transitions 
can be governed by local equilibrium thermodynamics or by active processes controlled by biological fuel.
It remains largely unexplored how the behavior of phase-separating systems with molecular transitions differs between thermodynamic equilibrium and cases where detailed balance of the molecular transition rates is broken due to the presence of fuel. 
Here, we present a model of a phase-separating ternary mixture where two components can convert into each other. 
At thermodynamic equilibrium, we find that molecular transitions can give rise to a lower dissolution temperature and thus reentrant phase behavior.
Moreover, we find a discontinuous thermodynamic phase transition in the composition of the droplet phase if both converting molecules attract themselves with similar interaction strength.
Breaking detailed-balance of the molecular transition leads to quasi-discontinuous changes in droplet composition by varying the fuel amount for a larger range of inter-molecular interactions.
Our findings showcase that phase separation with molecular transitions provides a versatile mechanism to control properties of intra-cellular and synthetic condensates via discontinuous switches in droplet composition.
\end{abstract}


\maketitle


\section{Introduction}

The ability to phase separate in aqueous solutions and to interconvert between distinct molecular states are two properties common among a large number of biomacromolecules. For example, DNA sequences composed of self-complementary strands phase-separate into DNA-rich condensates that coexist with a DNA-poor phase~\cite{Morasch2016, Nguyen2017, Jeon2018}.
These sequences can also switch between an expanded (``open'') DNA configuration and a collapsed (``closed'') configuration where the self-complementary strands of the same DNA are base-paired.
Another example are proteins that form phase-separated condensates \textit{in vitro}~\cite{Broide1991,Dumetz2008, Patel2015} 
and in living cells~\cite{Brangwynne2009, Banani2016}. Proteins attract each other, as well as other macromolecules such as RNA and DNA, for example, by sticky amino-acid patterns such as RGG repeats~\cite{ Wang2018, Franzmann2019, Martin2020} or charged domains~\cite{Nott2015, Brady2017}. These interaction sites are considered to drive protein phase separation but also mediate conformational transitions where certain domains of the protein rearrange~\cite{Boixet2020}. 

In living cells, phase separation of macromolecules such as proteins and RNA can also be actively controlled by biological fuel. Hydrolysis of ATP by kinase and phosphatase, for example, enables the regulation of the phosphorylation states of proteins. The number of phosphoryl groups attached to a protein, in turn, affects its net charge~\cite{Monahan2017} and may modify localized interaction sites along the sequence~\cite{Li2012}. Thus, the phosphorylation state of proteins strongly determines their interactions and thereby their propensity to form aggregates and phase-separated compartments~\cite{Aumiller2015, Carlson2020, Liu2020}. 

To effectively account for the impact of fuel on chemical reactions and phase separation,
various theoretical models were proposed with chemical reactions that are devoid of thermodynamic constraints and thus break detailed balance of the rates~\cite{Glotzer1994,  lee2018novel, tjhung2018cluster, Weber2019}. 
These systems give rise to novel non-equilibrium phenomena such as bubbly-phase separation~\cite{tjhung2018cluster} and shape instabilities leading to droplet division~\cite{Zwicker2017}.
Up to now, there has been a strong focus on establishing new universality classes and observing new non-equilibrium phenomena. 
However, the physical mechanisms of how transitions between different molecular states affect phase separation remain largely unexplored -- despite their ubiquity in biological systems. In particular, it remains unclear how these systems respond to thermodynamic perturbations such as temperature and concentration changes, and how fuel-driven molecular transitions affect the phase separation behavior. This is mainly due to the lack of physical models for phase separation and chemical reactions with coarse-grained parameters that are linked to molecular properties of biological macromolecules.

To gain insight into how molecular transitions are driven by fuel control phase separation, we propose a thermodynamic model for polymers, such as DNA and proteins, which can phase-separate and undergo molecular transitions between two molecular states. At thermodynamic equilibrium, we show that the interplay between phase separation and the molecular transitions gives rise to \emph{reentrant} phase behavior as a function of temperature.
Moreover, we also find a  discontinuous thermodynamic phase transition in droplet composition when changing the temperature.
Similar behavior also exists away from thermodynamic equilibrium where consumption of fuel breaks detailed balance of the rates associated with the molecular transition.
In this case,  droplet composition can abruptly switch upon small changes in fuel level.
In contrast to the corresponding thermodynamic system, the fuel-driven molecular transition releases the Gibbs phase rule constraint. This release allows abrupt switches in droplet composition
within a range of molecular interaction parameters that is significantly broader compared to the system at thermodynamic equilibrium. 
Owing to the very general nature of our model, our findings suggest that many different phase-separating macromolecules that are capable of undergoing molecular transitions
can switch the composition of dense phase via fuel-driven molecular transitions.

\section{Model for phase separation and  molecular transitions}

\begin{figure*}[t]
\centering
    \includegraphics[width=0.8\textwidth]{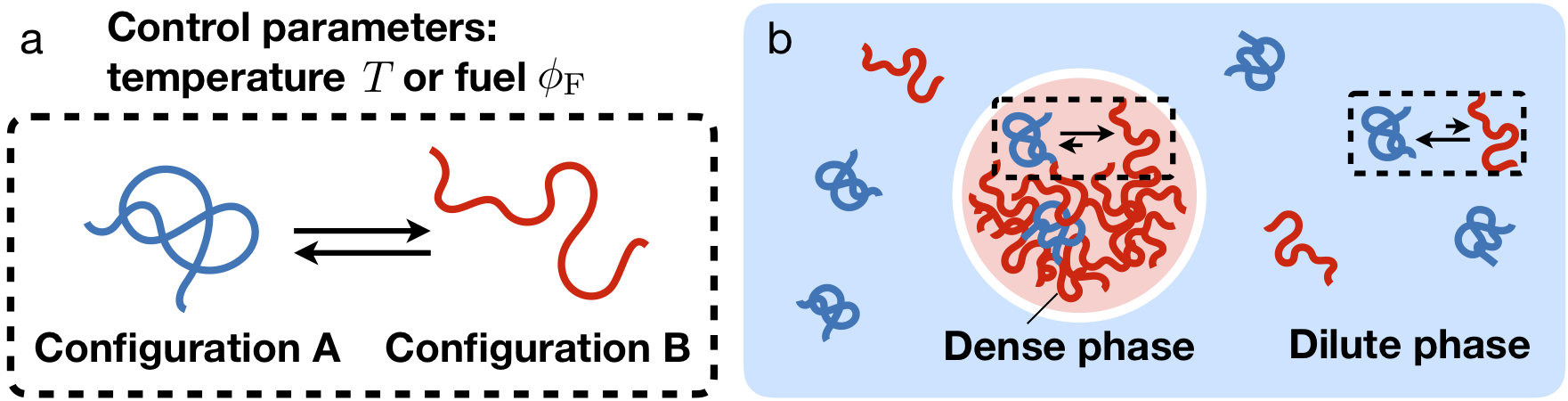}
    \caption{ 
    \textbf{Illustration of the model to study molecular transitions and phase separation.} In our model, we consider three components: $A$, $B$ and solvent $W$. (a) We consider reversible molecular transitions between $A$ and $B$ which are controlled by temperature $T$ or fuel volume fraction $\phi_\text{F}$. 
    (b) Molecular transitions occur in  the dense and dilute phase, respectively.  Depending on the molecular interactions of $A$ and $B$ molecules, the total volume fraction of both components and their  composition (i.e. the relative amount of $B$ molecules) change in each of the phases.
    }
    \label{fig:teaser}
\end{figure*}

Here, we consider mixtures composed of  unstructured macromolecules, such as DNA or proteins, and a solvent $W$.
In our model, these macromolecules can undergo molecular transitions (e.g. changes in conformation or charge) between two states, referred to as $A$ and $B$. The transition between the two states is described similarly to a chemical reaction with the scheme (Fig.~\ref{fig:teaser}a). The corresponding transition rates depend on thermodynamic control parameters such as temperature $T$ and macromolecules concentration, as well as fuel volume fraction $\phi_\text{F}$  that drives the transition away from thermodynamic equilibrium.
In our model, both components $A$ and $B$ can phase-separate due to attractive interactions among them 
(Fig.~\ref{fig:teaser}b).

%


\subsection{Free energy density}
We describe the mixture in the $T$-$V$-$N_i$-ensemble making use of a free energy density $f(T,\phi_i)=F(T,V,N_i)/V $, where $F$ denotes the Helmholtz free energy depending on temperature $T$, volume $V$  and particle number $N_i$ of component $i$.
Particle numbers are related to volume fractions $\phi_i= {N_i \nu_i}/{V}$
with $\nu_i$ denoting the corresponding molecular volumes.  To describe the thermodynamic behavior of unstructured polymers, we use a Flory-Huggins free energy density of the form
\begin{subequations}
\begin{equation}
\label{eq:f}
    f = 
    \frac{k_{B} T}{\nu_{W}} \left[
     \sum_i \frac{\phi_i}{r_i} \ln \phi_i + \sum_{i,j} \Omega_{ij} (T)\phi_i \phi_j+ \sum_i  \omega_i (T) \phi_i   \right]\, , 
\end{equation}
\newline
where $k_{B}$ is the Boltzmann constant, $r_i=\nu_i/\nu_{W}$ are the ratios of the molecular volumes relative to the solvent $W$, and $i=A,B,W$. For simplicity, we consider $A$ and $B$ having equal molecular volumes:  $\nu_{A} = \nu_{B} \equiv \nu$ and thus 
$r_{A} = r_{B} \equiv r$, implying that the molecular transition conserves the volume
of the system, $V=\sum \nu_i N_i$. Moreover, in our model, all molecular volumes are constant making the system incompressible at all times
and leading to the incompressibility condition  
\begin{equation}\label{eq:incompressibility}
\phi_{W}    = 1 - \phi_{A} - \phi_{B} \, .
\end{equation}
For simplicity, we consider the fuel (and potential waste products) to be dilute or having a small molecular volume, thus not affecting the incompressibility condition. 
The first term in Eq.~\eqref{eq:f} is the mixing entropy governing the tendency of the mixture to remain mixed~\cite{Flory1942,Huggins1942}. The second term describes the interactions among the molecules, where $\Omega_{ij}$ denotes the interaction parameter between the components $i$ and $j$. These interaction  parameters  depend on temperature. To lowest order, we can write~\cite{Rubinstein2003}  
\begin{equation}
\label{eq:eij}
\Omega_{ij} =   \frac{e_{ij} - s_{ij} T}{k_{B} T}  \, ,
\end{equation}
\end{subequations}
where $e_{ij}$ and  $s_{ij}$ are temperature independent interaction energies and entropies per molecule. 
The third term in Eq.~\eqref{eq:f} accounts for the free energy needed to create 
the molecular components (e.g. via molecular transitions) and 
$\omega_i(T)$ denotes the
internal free energy of component $i$ (in units of $k_{B}T$).
In particular, $\omega_{B} - \omega_{A}$ corresponds to the internal free energy difference between the two molecular states $A$ and $B$, which we express as 
\begin{equation}
\label{eq:int_f_e}
\omega_{B} - \omega_{A} = \frac{ e_\text{int} - s_\text{int} T }{k_{B} T}  \, .
\end{equation}
Here, $e_\text{int}$ and $ s_\text{int}$ are the energetic and entropic differences between state $A$ and $B$. As a prototypical example, we consider conformational transitions leading to an energy and
entropy increase converting $A$ to $B$, i.e. $e_\text{int}>0$ and $s_\text{int}>0$. Such a scenario corresponds for example to polymers switching from a tightly bound configuration to an unfolded one. 
For a dilute system, the temperature at which the internal free energy difference vanishes, $\omega_{B}(T) =\omega_{A}(T)$, defines the melting temperature
\begin{equation}\label{eq:T_melt}
    T_\text{m} = e_\text{int}/s_\text{int} \, ,
\end{equation}
above/below which $B$/$A$ is the favored molecular state in the dilute limit where the interactions terms in Eq.~\eqref{eq:f} are expanded up to the first order in volume fraction.

\subsection{Thermodynamic equilibrium}
\label{subsec:eq}

Chemical equilibrium is reached between molecular states if~\cite{Alberty2003book, Gong2004, Adame_Arana2020}
\begin{equation}
    \label{eq:chem_bal_ex}
    \bar{\mu}_{A} = \bar{\mu}_{B} \, ,
\end{equation}
where the exchange chemical potentials are given by~\cite{Gong2004}
\begin{equation}
    \label{eq:mu_ex}
    \bar{\mu}_i = \nu_i \frac{\partial f}{\partial \phi_i} \, .
\end{equation} 
Here, $f$ is the free energy density after using the incompressibility condition Eq.~\eqref{eq:incompressibility}. Exchange chemical potentials are related to the chemical potential differences difference between molecules and solvent, as explained in detail in Appendix~\ref{app:thermo_equ}. Using the free energy density $f$ (Eq.~\eqref{eq:f}), 
the condition of chemical equilibrium can be written as
\begin{equation}
\label{eq:chem_curve}
\frac{\phi_{B}}{\phi_{A} }=  \exp \left[ - r \left(\omega_{B} - \omega_{A} + 2 \sum_{i={A,B,W}} (\Omega_{Bi} - \Omega_{Ai}) \phi_i \right) \right] \, .
\end{equation}
This relationship represents the non-ideal mass action law for the molecular transition depicted in Fig.~\ref{fig:teaser}; for a general discussion of non-ideal chemical reactions in multi-component mixtures, see Ref.~\cite{GittermanBook}. The condition of chemical equilibrium~\eqref{eq:chem_curve}, together with incompressibility Eq.~\eqref{eq:incompressibility}, reduces the number of independent variables from three to one (see Appendices~\ref{sec:red_ind_var} and~\ref{sec:cheq_incompr} for details). 
Here, we consider 
the total volume fraction of $A$ and $B$
\begin{equation}\label{eq:phitot}
\phi_\text{tot}=\phi_{A} + \phi_{B} \, ,
\end{equation}
which is conserved in the molecular transition,
as independent variable.
By means of Eqs.~\eqref{eq:chem_curve} and \eqref{eq:phitot},
we can recast Eq.~\eqref{eq:f} in the form $f = f(T, \phi_\text{tot})$, which can be used to determine the phase diagram via the common tangent construction (i.e., Maxwell construction).  
In fact, imposing the equivalence between exchange chemical potential and osmotic pressure in both phases, the conditions for phase coexistence of the phase I and II read~\cite{Safran2003}:
\begin{subequations} \label{eq:Max_constr}
\begin{align}
\bar{\mu}_A(\phi_\text{tot}^\text{I}) &= \bar{\mu}_A(\phi_\text{tot}^\text{II}) \, , \\
\bar{\mu}_A  (\phi_\text{tot}^\text{I}) &= \nu_A \frac{ f(\phi_\text{tot}^\text{II})-f(\phi_\text{tot}^\text{I}) }{\phi_\text{tot}^\text{II}-\phi_\text{tot}^\text{I}} \, .
\end{align}
\end{subequations}
An example of Maxwell construction for the free energy density at chemical equilibrium is shown in Fig. \ref{fig:f_loop}, in Appendix~\ref{sec:f}.

\subsection{Non-equilibrium thermodynamics}
\label{sec:non_eq_thermo}

Using irreversible thermodynamics, the kinetic equations for an incompressible, ternary mixture read (see Appendix~\ref{app:non_eq} for the derivation)
\begin{subequations}\label{eq:phiAphiB}
\begin{align}
	\partial_t \phi_A &= \nabla \cdot \left( \Lambda_A \nabla \tilde{\mu}_A  \right) + s(\phi_A, \phi_B) \, ,\\
	\partial_t \phi_B &= \nabla \cdot \left( \Lambda_B \nabla \tilde{\mu}_B  \right) - s(\phi_A, \phi_B) \, ,
\end{align}
\end{subequations}
where the inhomogeneous chemical potentials are given as~\cite{Kruger2018}
\begin{equation} \label{eq:mu_bar}
	\tilde{\mu}_i= \bar{\mu}_i -\kappa_i \nabla^2 \phi_i {-} \kappa  \nabla^2 \phi_{j} \, .
\end{equation}
These chemical potentials are related to the free energy 
\begin{equation}
\label{eq:F_extensive}
 F = \int \text{d}^3x \, \bigg( f + \sum_{i=A,B} \frac{\kappa_i}{2 \, \nu } |\nabla \phi_i |^2  + \frac{\kappa}{\nu } \nabla \phi_A \cdot \nabla \phi_B \bigg) 
\end{equation}
via $\tilde{\mu}_i=\nu_i \delta F/\delta \phi_i$, where the free energy density $f$ is given by Eq.~\eqref{eq:f}.
Moreover, $\Lambda_i$ is the diffusive mobility which depends on volume fraction. To ensure that the diffusion coefficient of $A$ and $B$ is constant in the dilute limit, we use the following scaling form, $\Lambda_i = \Lambda_{i,0} \phi_i \left( 1- \phi_A -\phi_B \right)$~\cite{Langer1974}, and consider  $\Lambda_{i,0}$ to be constant for simplicity.
This choice indeed cancels the divergence stemming from the logarithmic terms in the free energy density.
Thus, the diffusion constant in the dilute limit is given by $D_i=k_{B}T \Lambda_{i,0}$. For simplicity,  we restrict ourselves to the special case of a zero Onsager cross-coupling coefficient. 
 
The kinetics of the chemical transition is captured by the reaction flux  $s(\phi_A, \phi_B)$ which depends on the volume fraction of both molecular components. 
Here, we consider two different cases corresponding to different reaction fluxes. 
The first case is a system that relaxes toward thermodynamic equilibrium, suggesting the following  form (for the derivation using linear response, see Appendix~\ref{app:non_eq}):
\begin{equation}\label{eq:s_to_EQ}
	s = {\Lambda}_s \frac{ \left( \tilde{\mu}_B - \tilde{\mu}_A \right)}{k_{B}T} \, ,
\end{equation}
where ${\Lambda}_s$ denotes the mobility for the molecular transition, which we consider to be constant for simplicity. If chemical potentials are homogeneous ($\tilde{\mu}_i=\text{const.}$) and the chemical potentials of $A$ and $B$ are equal ($\tilde{\mu}_A = \tilde{\mu}_B$), the system is at thermodynamic equilibrium, self-consistently leading to $\partial_t \phi_A=\partial_t \phi_B=0$ in Eqs.~\eqref{eq:phiAphiB}.

The second case refers to a system, where the molecular transitions cannot relax toward thermodynamic equilibrium, i.e., detailed balance of the rates is broken~\cite{Julicher1997, Weber2019}. 
In living or active systems, this is often facilitated by a ``fuel'' component F which affects the balance between 
the two molecular states and is, to a good approximation, maintained by chemical reaction cycles~\cite{tenaaccelerated, donau2020active}. 
Here, we consider a combination of the flux~\eqref{eq:s_to_EQ} and a second order chemical reaction that depends on the fuel volume fraction $\phi_\text{F}$ (see Appendix~\ref{app:brok_det_bal} for details):
\begin{align} \label{eq:s_fuel}
s & = \Lambda_s    \frac{ (\tilde{\mu}_{B}-\tilde{\mu}_{A}) }{k_{B}T}  + k_\leftarrow \phi_{B}\phi_\text{F} -k_\rightarrow \phi_{A}\phi_\text{F} \, .
\end{align}
Here,
$k_\leftarrow $ and $k_\rightarrow $ are the independent rate constants of the backward and forward transition, respectively. This independence of rate constants implies that detailed balance of the rates corresponding to the molecular transition is broken; for a conceptual discussion see Ref.~\cite{Weber2019}.
In contrast to Eq.~\eqref{eq:s_to_EQ}, 
a system with a reaction flux $s$ given by Eq.~\eqref{eq:s_fuel} cannot fulfill the two equilibrium conditions of equal and spatially constant chemical potentials. 
Thus, stationary solutions to Eqs.~\eqref{eq:phiAphiB} using Eq.~\eqref{eq:s_fuel} are non-equilibrium steady states. Consistently with this, in the absence of fuel ($\phi_\text{F}=0$), the reaction flux above reduces to Eq.~\eqref{eq:s_to_EQ} and the system can relax to thermodynamic equilibrium. In our model, the fuel level controls how far the system is maintained away from thermodynamic equilibrium.   

Finally, we look for an equation for the fuel volume fraction $\phi_\text{F}$. 
To this end, we focus on the case where diffusion of fuel is fast compared to diffusion of the macromolecules $A$ and $B$, respectively. This limit is indeed reasonable for many biological systems since diffusivities for example between phase-separating macromolecules (proteins, RNA,...) and ATP  differ by about two orders in magnitude~\cite{Nenninger2010,Hubley1995}.
For simplicity, we consider the case of fuel being conserved, i.e., it is maintained constant in time.
This scenario applies to living cells under physiological conditions and in \emph{in vitro} systems, where these conditions could be realized by encapsulated ATP or regeneration of ATP.
Moreover, we assume that the fuel molecules  interact in the same way with $A$ and $B$.
In this case, we can quasi-statically slave the fuel volume fraction $\phi_\text{F}$ to the total concentration of $A$ and $B$,
\begin{equation}\label{eq:phi_fuel}
	\phi_\text{F} (\vect{x},t) = \bar{ \phi}_\text{F} \left(\alpha  +  \beta \,\phi_\text{tot}(\vect{x},t) \right)\, ,
\end{equation}
where
$\bar{\phi}_\text{F}$ denotes the average volume fraction of fuel that is constant in time.
The choice above allows to capture the partitioning of fuel by accounting for the spatial correlations between 
fuel and the total volume fraction $\phi_\text{tot}$.
The fuel partitioning coefficient $P_\text{F}$, that is experimentally accessible, determines the values of the parameters $\alpha$ and $\beta$ in Eq.~\eqref{eq:phi_fuel} (see Appendix~\ref{app:fuel_eq_part} for a definition of $P_\text{F}$ and the link to $\alpha$ and $\beta$).

In the following, we choose three parameter sets for $\alpha$ and $\beta$ corresponding to three qualitatively different scenarios.  First, the fuel partitions inside the $\phi_\text{tot}$-rich phase for $\alpha =0$ and $ \beta=1/\bar{\phi}_\text{tot}$.
Second, fuel is enriched outside 
corresponding to $\alpha = -\beta = 1/(1-\bar{\phi}_\text{tot})$. And finally, we also consider 
the case of a homogeneous fuel  for $\alpha =1 , \beta=0$. The latter case has been studied for example in Refs.~\cite{Glotzer1994, Zwicker2015}.

\section{Equilibrium phase diagrams}

In this section, we study the equilibrium phase diagrams as a function of the total volume fraction $\phi_\text{tot}$ and a scaled temperature $T/T_0$, with  $T_0=-e_{BB}/k_\mathrm{B}$. In such phase diagrams, the binodal lines separate demixed and mixed thermodynamic states.
Along the binodals, we also depict the composition in terms of the relative abundance of $B$ molecules,  $\phi_B/\phi_\text{tot}$ (see the color code in Fig.~\ref{fig:examples}). We then study how such phase diagrams are affected by the melting temperature $T_\text{m}$.
As a reference system, we consider a binary mixture composed of only $B$ and $W$ molecules (black lines in Figs.~\ref{fig:examples}a-c). In all our studies, we choose $r=2$ to account for differences in molecular volumes between macromolecules and water.

\subsection{Reentrant phase behavior}

\begin{figure*}[t]
\centering
    \includegraphics[width=1.0\textwidth]{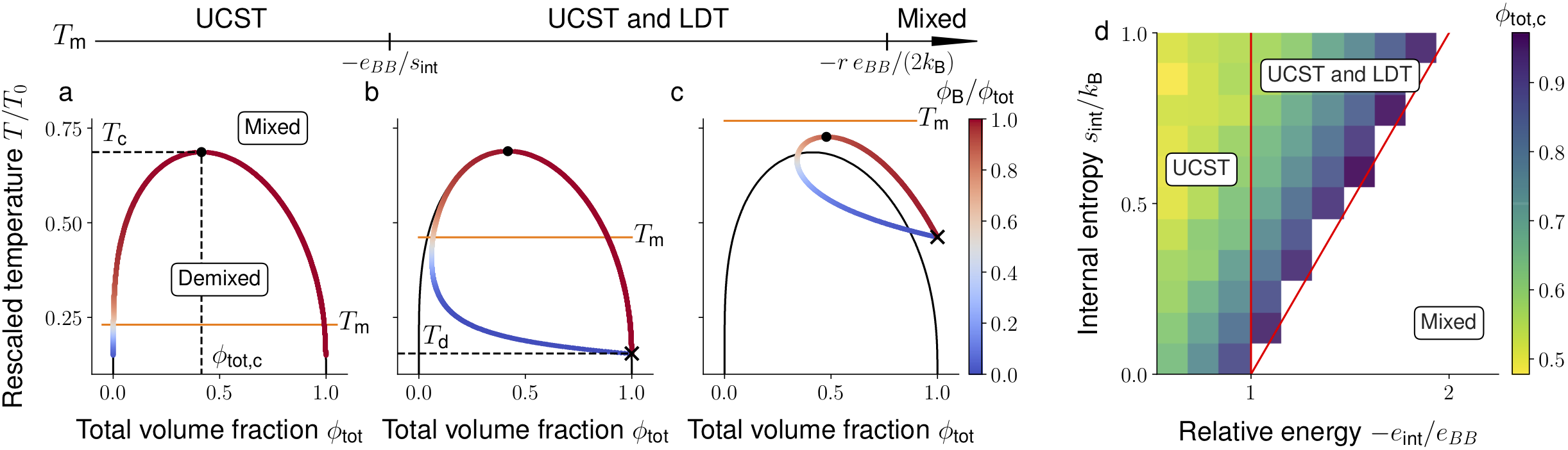}
    \caption{ 
    \textbf{Molecular transitions can lead to a lower critical dissolution temperature.}
    The phase diagrams depict the mixed and demixed region in the plane spanned by total volume fraction $\phi_\text{tot}$ and rescaled temperature $T/T_0$. The composition of each phase is indicated by the color code displayed on each branch of the binodal. The black line corresponds to the binary $B$-$W$ mixture. For increasing $T_\text{m}$, the presence of  molecular transitions reshape the phase diagram.
    (a) For low $T_\text{m}$, the demixed region is bound from above by an upper critical solution temperature (UCST) similar to the binary case. 
    (b) For $T_\text{m} > -e_{BB}/s_\text{int}$, the phase diagram becomes {reentrant}, i.e., bounded also from below by a lower dissolution temperature (LDT). (c) Increasing $T_\text{m}$ further, the binodal is upshifted until the demixing region shrinks into a point for $T_\text{m} =  - r \, e_{BB}/(2 k_\text{B}) $. Parameters for (a-c): $-e_\text{int}/e_{BB} = 0.75, 1.5, 2.5$
    (d) Influence $e_\text{int}$ and $s_\text{int}$ on the phase behaviour, where the color code represents the critical density.  Increasing the interaction entropy $s_\text{int}$, the interval of relative energies -$e_\text{int}/e_{BB}$ corresponding to the reentrant regime widens. All the results presented correspond to $r=2$. 
    }
    \label{fig:examples}
\end{figure*}

We first study the case of only attractive homotypic $B$-$B$ interactions and neglect entropic contributions for simplicity  ($s_{ij}=0$ in Eq.~\eqref{eq:eij}). Specifically, $\Omega_{BB} =  e_{BB}$ and $ \Omega_{ij} = 0$ otherwise.
Attractive homotypic interactions lead to phase coexistence between a $\phi_\text{tot}$-rich and a  $\phi_\text{tot}$-poor phase, in the following referred to as the dense and dilute phase, respectively.   
Below the melting temperature $T_\text{m}$ (Eq.~\eqref{eq:T_melt}),
we find that coexisting phases are not only different in $\phi_\text{tot}$, but can also differ in the amount of $B$ relative to $A$ (see color code in Fig.~\ref{fig:examples}a).
In particular, the dense branch of the binodal corresponding to large values of $\phi_\text{tot}$ is rich in $B$ components, while the composition of the dilute branch (low $\phi_\text{tot}$) changes with temperature. 
The composition, defined as the relative amount of molecules in the $B$ conformation, $\phi_B/\phi_\text{tot}$, remains rather uniform in the dense branch. In fact, in this example, attraction among $B$ molecules is the only interaction considered, thus the dense phase must be composed dominantly of $B$ molecules.
However, in the dilute and $B$-poor phase, the fraction of $B$ is affected by temperature since the molecular transition, which favors $A$ over $B$ at low temperatures, dominates over the homotypic $B$-$B$ interaction.

Increasing the melting temperature $T_\text{m}$
enhances the dominance of the molecular transition at low temperatures leading to a qualitative change of the thermodynamic phase diagram (Fig.~\ref{fig:examples}b). In fact, we find a lower dissolution temperature (LDT), which implies a \textit{reentrant} phase behavior. Such behavior is manifested as the possibility that both increasing and decreasing the temperature leads to a phase transition from a demixed to a mixed state, respectively. 
The lower dissolution temperature is given by (for a derivation see Appendix~\ref{app:trans_point})
\begin{equation}\label{eq:Td}
    T_\text{d}=T_\text{m} +  {e_{BB}}/{s_\text{int}} \, ,     
\end{equation} 
and is set by the competition between the homotypic interactions $e_{BB}$ and the molecular transition which is characterized by the melting temperature  $T_\text{m}$ (Eq.~\eqref{eq:T_melt}). For attractive interactions, Eq.~\eqref{eq:Td} fulfills $T_\text{d}<T_\text{m}$, which is consistent with the fact that 
the majority of molecules have to be in the $A$ state to undergo a phase transition to a mixed state for decreasing  temperatures.

For even higher melting temperatures  $T_\text{m}$,
the upper critical solution temperature $T_\text{c}$ (UCST) and critical volume fraction $\phi_\text{c}$ exceed the corresponding values in the binary case (black line in Fig.~\ref{fig:examples}a-c).
This is because below $T_\text{m}$, the dilute and dense phases differ in composition, being enriched in $A$ and $B$, respectively. If $T_\text{m}$ is larger than the critical temperature of the reference binary system, this composition difference will emerge as soon as the homogeneous system demixes into two phases.  
Even though, with our choice of interaction parameters,
$B$ effectively repels both $A$ and the solvent, differences in molecular volumes of $A$ and $B$ with respect to the solvent (i.e. $r>1$) favors phase-separating $B$ from $A$.
In other words,  two coexisting phases of different solvent content are entropically disfavoured. 
This entropic disadvantage implies that both phases will be up-shifted in $\phi_\text{tot}$ leading to two coexisting phases rich in $B$ and $A$, respectively, instead of $B$ and solvent rich, as in the binary reference. This entropic disadvantage of phase-separating $A$ with respect to $B$ also explains the increase in critical temperature.

The qualitative difference among the phase diagrams can be summarized in terms of the internal entropy and energy differences between the two states $A$ and $B$ (Eq.~\eqref{eq:int_f_e}); see Fig.~\ref{fig:examples}d.
For $e_\text{int} < -e_{BB}$, the lower dissolution temperature  (Eq.~\eqref{eq:Td}) vanishes, leading to phase diagrams with only an upper critical solution temperature (UCST). This UCST arises from a competition between $B$-$B$ interactions and the mixing entropy (Fig.~\ref{fig:examples}a).
If the internal energy gain of the molecular transition exceeds the strength of attractive $B$-$B$ interaction, i.e., $e_\text{int} \gtrsim -e_{BB}$ (corresponding to $T_\text{m} \gtrsim -e_{BB}/s_\text{int}$), there is a finite lower dissolution temperature $T_\text{d}$ leading to \emph{reentrant} phase behavior. 
Reentrance arises from a competition between the attractive interactions $e_{BB}$ and the energetic difference between the two molecular states $e_\text{int}$ (see by inserting Eq.~\eqref{eq:T_melt} into Eq.~\ref{eq:Td}).
Reentrance is absent at low  $T_\text{m}$ because  $B$-$B$ interactions dominate the molecular transition ($e_\text{int} < -e_{BB}$). 
Increasing $T_\text{m}$ corresponding to 
$e_\text{int}>-e_{BB}$ 
causes reentrant behavior because the energetic difference between the two molecular states $e_\text{int}$ exceeds the strength of the attractive $B$-$B$ interaction. 
A further increase of $T_\text{m}$
leads to a transition to a mixed region in the phase diagram.
We found that the disappearance of demixed states in the phase diagram coincides with the upper critical point $(\phi_\text{c}, T_\text{c})$ reaching $\phi_\text{tot}=1$, which corresponds to a system without solvent (see color coded $\phi_\text{c}$ in Fig.~\ref{fig:examples}d).
This observation can be used to analytically calculate the transition line between reentrant phase behavior (UCST and LDT) and mixed states 
by $T_c(\phi_\text{tot}=1)
=T_\text{d}$, where $T_c(\phi_\text{tot}=1) =   -r \, e_{BB}/(2k_{B})$
(for details see Appendix~\ref{app:trans_point}). We find a relationship  for the transition line,  $-e_\text{int}/e_{BB} = 1 +  s_\text{int} \,{r}/{2}$, 
which agrees with results from the Maxwell construction depicted in Fig.~\ref{fig:examples}d.
This relationship implies that the 
reentrant region widens with the internal entropy difference  $s_\text{int}$.
Larger $s_\text{int}$ leads to more phase-separating $B$ molecules  and thus favors phase coexistence over mixing. 

\begin{figure}
    \includegraphics[width=\textwidth]{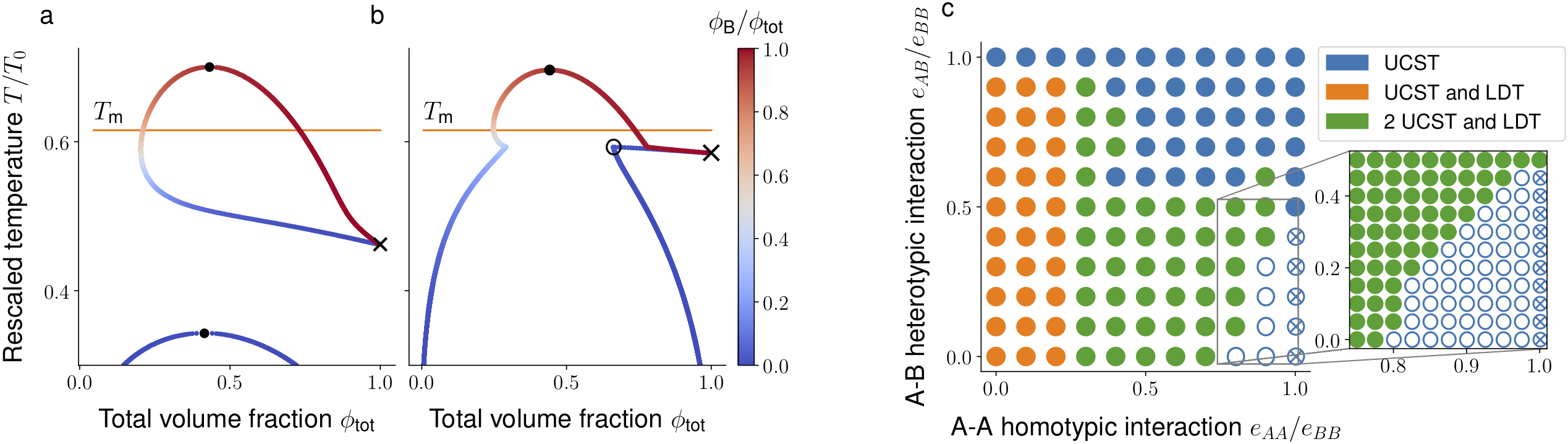}
    \caption{
    \textbf{Interplay between homo- and heterotypic interactions determine the phase diagrams of phase-separating systems with molecular transitions. }
    (a,b) $\phi_\text{tot}$ - $T/T_0$ phase diagrams with the composition along the binodal color coded, depicted for two prototypical cases discussed in the paper. (a) A weak self interaction among the $A$ components 
    creates a demixed region at low temperature which leads to a second  upper critical solution temperature (UCST). Here, $e_{AA} = 0.5\, e_{BB}$ and $e_{AB} = 0$. 
    (b) Increasing the strength of $A$-$A$ homotypic interaction
    let the lower demixed region merge with the upper reentrant region which generates a triple point. Here $e_{AA} = 0.9 \, e_{BB}$ and $e_{AB} = 0.1 \,e_{BB}$. 
    (c) Classification of phase diagrams for different heteropypic $A$-$B$ and homotypic $A$-$A$ interactions, keeping $e_{BB}$ and $T_m$ fixed. The classes are: phase diagrams with one UCST, phase diagrams with reentrant behavior which either have one UCST and one LDT or  two UCSTs and one LDT. The $\Circle$ symbol marks phase diagrams with a triple point (as shown in (b)) while $\otimes$ highlights cases in which the triple point collapses giving rise to a jump in the composition of the dense phase (see Fig.~\ref{fig:ph_trans}). Please note that for this classification, we considered temperatures down to
    $T/T_0=0.1$. Here and for the rest of the paper, we fix $T_m=0.615 \, T_0$.}
    \label{fig:interactions}
\end{figure}

\subsection{
First order phase transition in droplet composition}

Here we discuss the case of attractive homotypic interactions for both molecular states, i.e., $e_{AA}<0$ and $e_{BB}<0$ and quantify the impact on the phase diagram of the relative $A$-$A$ interaction strength $e_{AA}/e_{BB}$.
For simplicity, we now  fix $T_\text{m}$ to a value for which reentrant behavior was observed in the case of only $B$-$B$ interactions, as discussed in the last section.
Due to the attractive homotypic $A$-$A$ interaction, we find 
an additional demixing region in the phase diagram at low temperatures. This region is disconnected from the reentrant region of the phase diagram, which is located at higher temperatures (Fig.~\ref{fig:interactions}a).
This new region at low temperatures has an upper critical solution temperature (UCST) and describes phase coexistence between  $\phi_\text{tot}$-rich and $\phi_\text{tot}$-poor phases that are both mainly composed of $A$. Thus, these phase diagrams have two UCSTs and one LDT. 

The phase diagram qualitatively changes when the strength of $A$-$A$ attraction is increased towards the value of $B$-$B$ interaction strength. In this case, the two separated domains in the phase diagram in Fig.~\ref{fig:interactions}a merge.
The resulting phase diagrams exhibit either only one UCST (not shown) or  
one UCST and a triple point; see Fig.~\ref{fig:interactions}b.
At this triple point, three phases coexist which differ in $\phi_\text{tot}$ and $A$-$B$ composition. 
The precise value of $e_{AA}$ at which both domains merge is influenced by the strength of the $A$-$B$ heterotypic interaction; see Fig.~\ref{fig:interactions}c.

\begin{figure}
    \includegraphics[width=\textwidth]{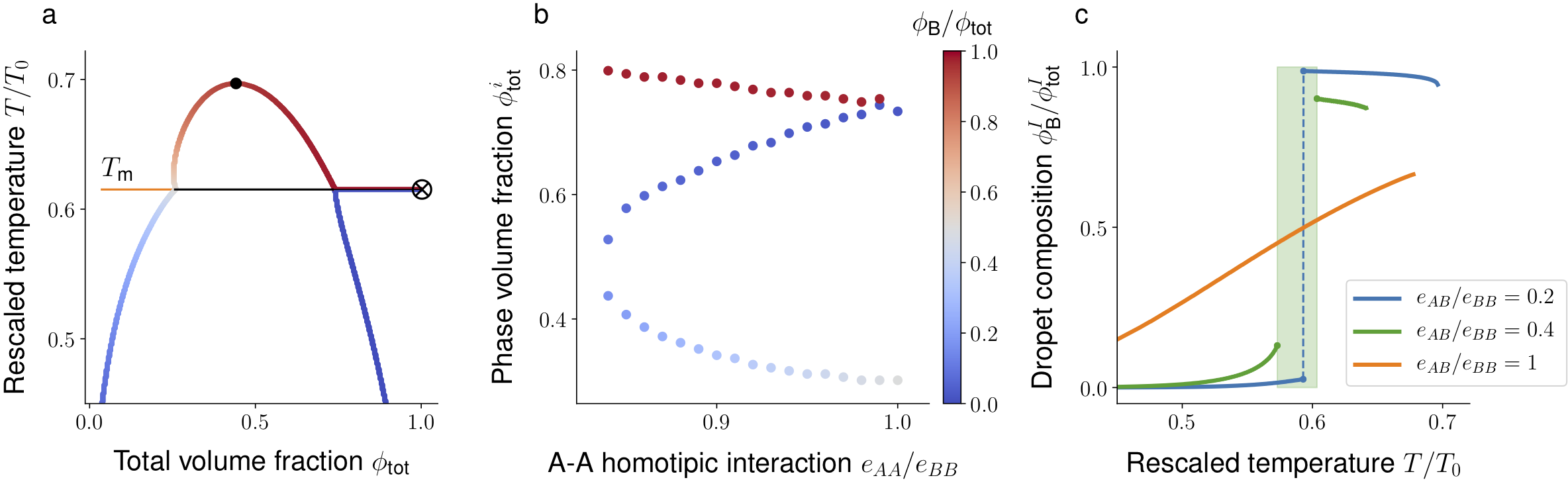}
    \caption{
    \textbf{Systems with strong homotypic interactions exhibit a first order phase transition in the composition of thre dense phase.}
    (a) In the limiting case of identical homotypic interaction strength combined with weak heterotypic interactions, $e_{AA} = e_{BB}$, $ e_{AB}/e_{BB} \ll 1$, the triple point collapses into a point at which the composition of the dense phase changes discontinuously. (b) Volume fraction of the three phases coexisting at the triple point as a function of relative $A$-$A$ interaction strength, $e_{AA}/e_{BB}$. For equal homotypic interactions, i.e. $e_{AA}=e_{BB}$, the two upper branches merge giving rise to the discontinuous transition in droplet composition. 
    Here, $e_{AB}/e_{BB}$=0.2 and the color code depicts composition of the phases. (c) Increasing the heterotypic interaction strength, 
    the droplet composition shows three distinct behaviors with  temperature: it can change discontinuously ($e_{AB}/ e_{BB} = 0.2$), or the droplet first dissolves and then reforms with different composition ($e_{AB}/e_{BB} =  0.4 $), where the shaded region corresponds to no drops, or its composition changes gradually ($e_{AB}/e_{BB} =  1 $). We used $\phi_\text{tot} = 0.5$ for all the three curves.}
    \label{fig:ph_trans}
\end{figure}

For $e_{AA}$ being equal to the value $e_{BB}$, the triple point temperature coincides with both the melting temperature $T_\text{m}$ and the temperature of the first order transition point where the two branches of different $A$-$B$ composition meet (indicated by a cross in Fig.~\ref{fig:ph_trans}a).
The result is a triple line; see the black horizontal line in Fig.~\ref{fig:ph_trans}a.
At the corresponding triple line temperature,
three phases coexist for any value of $\phi_\text{tot}$ between the binodals.
Crossing the triple line temperature leads to a discontinuous phase transition of the composition of the dense phase. In other words, in a finite system with a single droplet, the composition of the droplet discontinuously switches upon temperature changes between  $A$-rich and $B$-rich. 
As mentioned, the triple line temperature is exactly the melting temperature  $T_\text{m}$ (defined in Eq.~\eqref{eq:T_melt}).
The reason is that for $e_{AA} = e_{BB}$, both molecular states have the same phase separation propensity leaving it to the internal free energy balance to determine the composition of the dense phase.  
For total volume fractions larger than the dense binodal branch, there is also a discontinuous switch in $A$-$B$ composition (see the black line in Fig.~\ref{fig:ph_trans}a), however, the system is homogeneous in this case.  
When $A$-$A$ and $B$-$B$ attraction strengths approach each other, the total volume fraction of the dilute branch decreases, and the two denser branches, each corresponding to different $A$-$B$ composition, merge (Fig.~\ref{fig:ph_trans}b).
Thus, also for the $A$-$A$ attraction slightly weaker than the $B$-$B$ attraction, crossing the temperature corresponding to the triple point leads to a jump in both  composition and total volume fraction of the dense phase. For  $e_{AA} = e_{BB}$, the jump only occurs in  composition while the total volume fraction changes continuously.  

The discontinuous change in the composition of the dense phase is tied to the existence of a triple point. This triple point in turn arises from the tendency of the ternary mixture to form three coexisting phases stemming from similar attractive interactions among $A$ and $B$ molecules, respectively.
However, molecular transitions affect the dimensionality of the three phase coexistence domain in the phase diagram.
Three phase coexistence is consistent with the Gibbs phase rule.
In fact, due to the molecular transition, the system is reduced to an effective binary mixture solely characterized by the conserved variable $\phi_\text{tot}$. 
This reduction does not suppress the three phase coexistence region since the Gibbs rule for a binary mixture at fixed temperature and pressure allows for a maximum of three coexisting phases. However, in our case, the molecular transition at thermodynamic equilibrium reduces the dimensionality of this region by at least one, giving rise to a triple point or even a triple line.

The coexistence of three phases and thus also the triple point is controlled by heterotypic interactions $e_{AB}$ (see Fig.~\ref{fig:ph_trans}c). In particular, for strong attractive $A$-$B$ interactions, three phase coexistence is suppressed. Consistently, the triple point vanishes for increasing heterotypic interaction strength 
implying that the composition of the dense phase be cannot  upon temperature changes.  Instead, reentrance forces phase separation to vanish within a narrow intermediate temperature window (green shaded domain in Fig.~\ref{fig:ph_trans}c).
For even larger heterotypic interaction strength,
the composition of the dense phase changes continuously (orange line in Fig.~\ref{fig:ph_trans}c).

\section{Kinetics of phase separation with detailed-balance broken molecular transitions}

In the last section, we focused on how temperature affects 
the composition of coexisting phases. 
Up to now, there is no evidence that living systems control their temperature to regulate phase separation. Thus, we discuss how fuel-driven molecular transitions that are maintained away from thermodynamic equilibrium can control the composition inside droplets. 
For example, phosphorylation involving the hydrolysis of ATP is known to drive molecular transitions and thereby regulate protein phase separation~\cite{Banani2016, Carlson2020, Liu2020}.

To this end, we consider two prototypical parameter sets: (i) weak homotypic interactions of the $A$ species and weak heterotypic $A$-$B$ interactions: $e_{AA}/e_{BB} \ll 1$, $ e_{AB}/e_{BB} \ll 1$  and (ii) strong attractive $A$-$A$ interactions and weak $A$-$B$ heterotypic interactions:  $e_{AA}= e_{BB}$, $ e_{AB}/e_{BB}\ll 1$. At thermodynamic equilibrium, we observed reentrant phase behavior for (i) and the discontinuous phase transition in droplet composition for (ii).
To account for the effects of fuel on the molecular transition, we consider both a thermodynamic and a fuel-related contribution to the reaction flux; see Eq.~\eqref{eq:s_fuel}.  Most importantly, the fuel-related contribution breaks detailed balance of the rates. We numerically solve the kinetic Eqs.~\eqref{eq:phiAphiB} and \eqref{eq:s_fuel} in two dimensions with periodic boundary conditions combining the energy quadratization method~\cite{YANG2017104, Review_EQ_2018, ZHAO2019382} with with the stabilization method~\cite{Shen2010} (see Appendix~\ref{app:kin_and_numerics} for details).

\begin{figure*}
\centering
 \includegraphics[width=\linewidth]{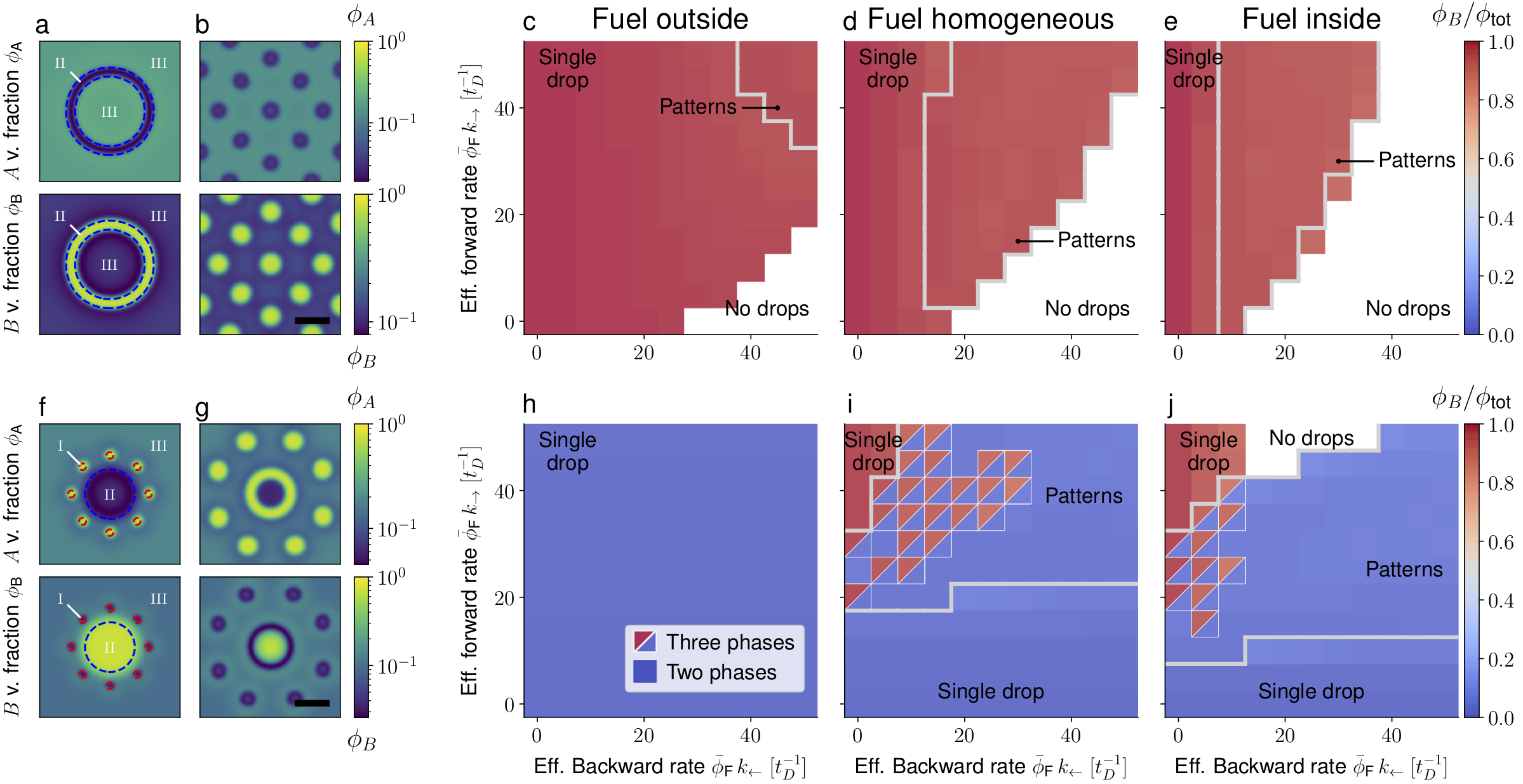}
\caption{
\textbf{Molecular transition breaking detailed balance leads to non-equilibrium stationary states.}
We monitor the emergence of patterns by tracking the interface between $A$-rich (I), $B$-rich (II), and solvent rich (III) domains (definition in App.~\ref{app:kin_and_numerics}). As an example, such tracked interfaces are shown by dashed line in (a) and (f), where the red and blue lines enclose $A$-rich and $B$-rich domains, respectively. These domains are the analogue of demixed phases at thermodynamic equilibrium. For weak $A$-$A$ self-interactions ($e_{AA} = 0.2 \, e_{BB}$, $ e_{AB} = 0 $), we display two prototypical non-equilibrium stationary states due to the presence of the fuel (a,b) and we determine the values of the effective rates corresponding to such out of equilibrium patterns (c)-(e), where the color code indicates the composition of the dense domain in terms of the relative $B$ amount, i.e., $\phi_{B}/\phi_\text{tot}$. Strikingly, fuel partitioning affects the boundaries of the region corresponding to patterns, along with the dissolution boundary. Here, $ \phi_\text{tot} = 0.35$, $T/T_0 = 0.55$. For strong $A$-$A$ self interactions ($e_{AA} = e_{BB}$, $ e_{AB} = 0.2 \, e_{BB} $), stationary patterns emerge that are composed of three domains distinct in composition (f) (g). In the diagrams (i)-(j), such domains are indicated with two triangles of different colors in which each color corresponds to the composition of the two distinct dense phases. Here $ \phi_\text{tot} = 0.35$, $ T/T_0 = 0.525$. The rates are measured in units of the inverse diffusion time $t_D = L^2 / D_A $, with $L = 100 \sqrt{\kappa_A/(k_\mathrm{B}T) }$ (see App. \ref{app:kin_and_numerics} for details). Scale bar: $20\sqrt{\kappa_A/(k_\mathrm{B}T)}$. }  
\label{fig:k_scan}
\end{figure*}

\subsection{Fuel controls non-equilibrium patterns}

Breaking detailed balance of the rates allows the system to settle into non-equilibrium stationary states which differ from thermodynamic equilibrium in terms of droplet composition and droplet number density. 
To classify these non-equilibrium states, we numerically solve Eqs.~\eqref{eq:phiAphiB} using Eq.~\eqref{eq:s_fuel} and initialize the system with one or two droplets of total volume fractions and $A$-$B$-composition corresponding to the thermodynamic phase diagrams. 
We find the emergence of various patterns ranging from equally sized droplets to rings and stripes (Fig.~\ref{fig:k_scan} (a), (b), (f), (g), and SI movies I-VII). The emergence of these patterns is determined by the partitioning of fuel into the droplets and the strength of  homotypic and heterotypic interaction of $A$ and $B$ molecules, respectively.

For the case (i) characterized by weak $A$-$A$ homotypic interactions, i.e. $e_{AA}/e_{BB} \ll 1$ and $ e_{AB}/e_{BB}<1$, we find that the $\phi_\text{tot}$-rich phase is mainly composed of $B$ and that the inside composition hardly varies due to the presence of fuel; Fig.~\ref{fig:k_scan} (c)-(e). 
However, the fuel can induce the formation of patterns that are significantly different from the initial single droplet at thermodynamic equilibrium (no fuel). 
For certain fuel partitioning, we find that there is an extended region in the $k_\rightarrow$-$k_\leftarrow$ state diagrams where patterns emerge (gray line in Fig.~\ref{fig:k_scan} (c)-(e)). Strikingly, the location and the extent of these regions  are significantly influenced by fuel partitioning. In particular,
fuel partitioning inside $\phi_\text{tot}$-dense domains favors the emergence of patterns for lower values of both effective forward and backward rate.
Close to the onset of pattern formation, we observe stable ring-like patterns (Fig.~\ref{fig:k_scan} (a) and SI Movie I). 

For increasing effective backward rate $\bar{\phi}_\text{F} \, k_\leftarrow$, we often observe 
 rings coexisting with droplets, see SI Movie II. 
For even larger effective rates $\bar{\phi}_\text{F} \, k_\leftarrow$, rings break-up leading to the emergence of equally sized droplets~(Fig.~\ref{fig:k_scan} (b) and SI Movie III). Equally sized droplets have been reported in a model using an  Ginzburg-Landau type of free energy~\cite{Zwicker2015}. 
For fuel partitioning inside, droplets tend to strongly deviate from their spherical shape (see SI Movie IV) and deform into elongated domains reminiscent of stripes in pattern-forming reaction diffusion system~\cite{frey2018protein}. Fuel partitioning also affects the boundary in the state diagram beyond which droplets dissolve.

For the case (ii) characterized by strong $A$-$A$ homotypic interactions,  i.e. $e_{BB}=e_{AA}$, $ e_{AB}/e_{BB}\ll 1$, the behavior in the $k_\rightarrow$-$k_\leftarrow$ state diagram changes significantly. 
For fuel partitioning outside, the composition is approximately independent of the rate constants and we exclusively find single, $A$-rich droplets that coexist with a solvent-rich phase (Fig.~\ref{fig:k_scan}h). 
The rate-independence implies that breaking detailed balance of the rates for the molecular transition does not enable control of phase separation in this setting. 
This changes if fuel is homogeneously distributed between the inside and outside or if fuel dominantly partitions inside the droplet phase (Fig.~\ref{fig:k_scan}  i,j). In these cases, increasing the effective forward rate for low effective backward rates leads to a transition from an $A$-rich droplet to a $B$-rich droplet.
In other words, for a large effective forward rate, this leads to a droplet of switched composition.

Between the two regions in the $k_\rightarrow$-$k_\leftarrow$ plane corresponding to single droplets of different composition (see Fig.~\ref{fig:k_scan} (i),(j)), there is a region where observe various kinds of patterns.
In particular, for a low effective forward rate corresponding to the onset of pattern formation, we typically find that
the shape of the $A$-rich domain deviates significantly from the initial single drop at thermodynamic equilibrium, while the composition of such domains remains close to the equilibrium value. 
Besides rings and equally sized drops, we also find patterns reminiscent of bubbly-phase separation~\cite{tjhung2018cluster}, see SI Movie V. 
For higher effective forward rates and low backward rates, we find patterns where three domains of different compositions stably coexist (see two triangles of different colors  in Fig. \ref{fig:k_scan} (i)-(j)).
Representative stationary patterns composed of three domains are shown in Fig.~\ref{fig:k_scan} f,g, and SI Movies VI - VII. 
These patterns are spherically symmetric with a centered droplet enriched in $B$ that is surrounded by a ring of smaller $A$-rich droplets of equal size. 
As mentioned above, for even higher effective forward rates, the system settles in a non-equilibrium stationary state composed of a single $B$-rich droplet.

\begin{figure}
\centering
\includegraphics[width=\linewidth]{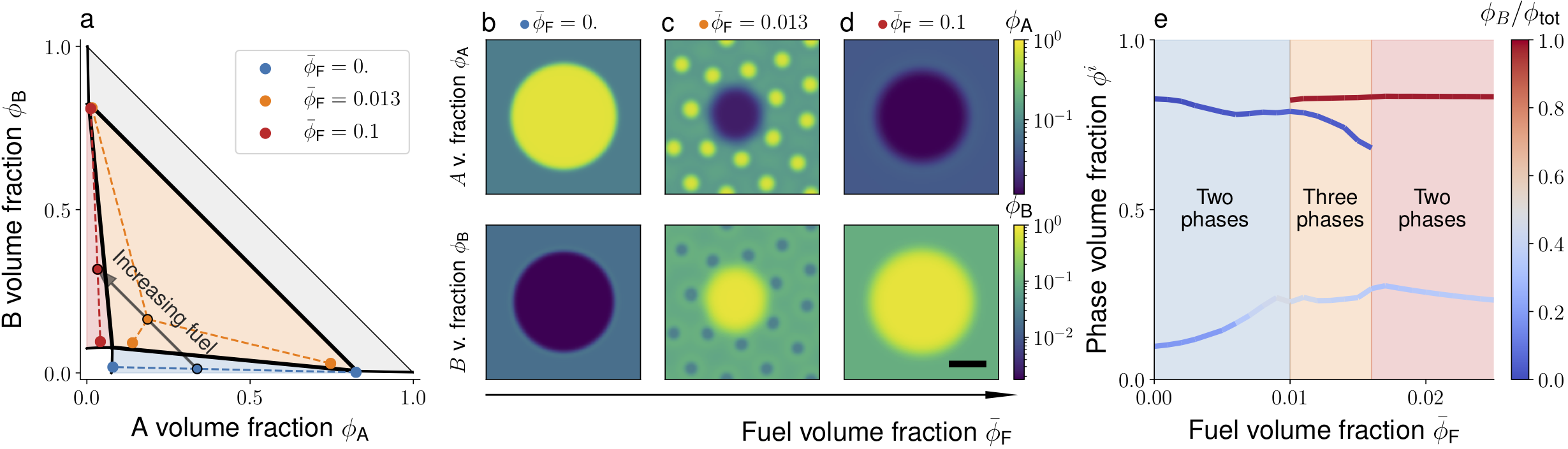}
\caption{ \textbf{Breaking detailed balance of the molecular transition enables controlling the number of distinct phases}
(a) Changing fuel in the systems can be illustrated as a trajectory in the ternary phase diagram. Without fuel the system demixes into two phases whose composition is set by thermodynamic equilibrium; the corresponding average volume fraction is indicated by the circled blue dot. The dashed lines connect the average volume fraction with the values in the respective demixed domains (blue dots). 
Increasing fuel pushes the average concentration from a region of two phase coexistence ($A$ and solvent rich, respectively) into a domain where three phase coexist at equilibrium (orange shaded triangle). Increasing fuel further leads to two domains,  
$A$ and solvent rich, respectively. 
Interestingly, the system demixes into domains whose volume fractions are well approximated by the corresponding equilibrium values. 
b)-d) $\phi_{A}$ and $\phi_{B}$ spatial profiles at the stationary state for increasing fuel average volume fraction $\bar{\phi}_\text{F}$. For moderate $\bar{\phi}_\text{F}$ the system exhibits three phase coexistence (b), while for high values the composition of the dense phase switches (c). (e) The number of phases along with their density and composition as a function of the amount of fuel in the system. The parameters are the same as the lower panel of Fig. \ref{fig:k_scan}, with fuel partitioning inside. Scale bar: $20\sqrt{\kappa_A/(k_\mathrm{B}T)}$.}
\label{fig:three_p}
\end{figure}

For high values of $k_\rightarrow$ and $k_\leftarrow$ in Fig.\ref{fig:k_scan} (h-j) and in contrast to (c-e), we find that the local composition of each domain is no more governed
by the equilibrium values
of a ternary mixture undergoing molecular transitions at thermodynamic equilibrium (as discussed in Sect.~\ref{subsec:eq}).
The reason is that the fuel-mediated molecular transitions (detailed balance broken)  dominate the thermodynamic ones
and that $A$-$A$ interactions are sufficiently strong.
Surprisingly, 
the local composition of each domain in the fueled system is almost equal to the equilibrium value
of a ternary mixture without molecular transitions. 
Contrary to the equilibrium system with molecular transitions that is effectively described by only one independent variable (i.e., $\phi_\text{tot}$),  a ternary mixture without transitions has two degrees of freedom (i.e., $\phi_A$ and $\phi_B$); the corresponding phase diagram for a fixed temperature is depicted in Fig.~\ref{fig:three_p}a. 
For ternary mixtures at thermodynamic equilibrium and parameters corresponding to the case (ii), there is a triangle in the $\phi_A$-$\phi_B$ phase diagram where the vertices of the triangle correspond to the composition of each of the coexisting phases at thermodynamic equilibrium (shaded in orange in Fig.~\ref{fig:three_p}a).
In the presence of the detailed balance broken molecular transition, the composition in each domain is actually close to the vertices of the equilibrium triangle (see orange circles in Fig.~\ref{fig:three_p}a). Changing the effective forward/backward transition rates amounts to moving the vector corresponding to the average volume fraction along the conserved trajectory $\phi_B=\phi_\text{tot}-\phi_A$. We now set the effective backward rate $ \bar{\phi}_\text{F} \, k_\leftarrow = 0$ and $k_\rightarrow = 2 \, 10^3 \,t_D^{-1}$ and consider variations of the effective forward rate induced by the average fuel amount. Specifically, in the absence of fuel ($\bar{\phi}_\text{F}=0$), an $A$-rich phase coexists with a solvent-rich phase at thermodynamic equilibrium (blue domain in phase diagram). Intermediate amounts of fuel induce a transition to a non-equilibrium stationary state comprised of three coexisting domains, see Fig.~\ref{fig:three_p}b-d for representative stationary states at different fuel levels. 
For an even larger amount of fuel, the system transits to a non-equilibrium stationary state where $B$-rich domains stably coexist with a solvent-rich domain. Although the three phase coexistence regime occupies the majority of the $\phi_A$-$\phi_B$-plane, coexistence of three domains is actually only accessible over a narrow range of fuel values. 
To see this, compare the orange regions along the conserved trajectory in Fig.~\ref{fig:three_p}a with the orange regions in Fig.~\ref{fig:three_p}e. Thus, when the fuel is changed by an amount larger than this narrow window, the long-time, stationary state of the systems swaps from an $A$-rich to a $B$-rich domain.
In general, this does not necessarily imply that the initial single droplet swaps its composition without dissolving. 
However, our results suggest the intriguing possibility, that under some conditions, a droplet may be able to change its composition without losing its identity related to the profile of the total volume fraction.

\begin{figure}
\centering
\includegraphics[width=\linewidth]{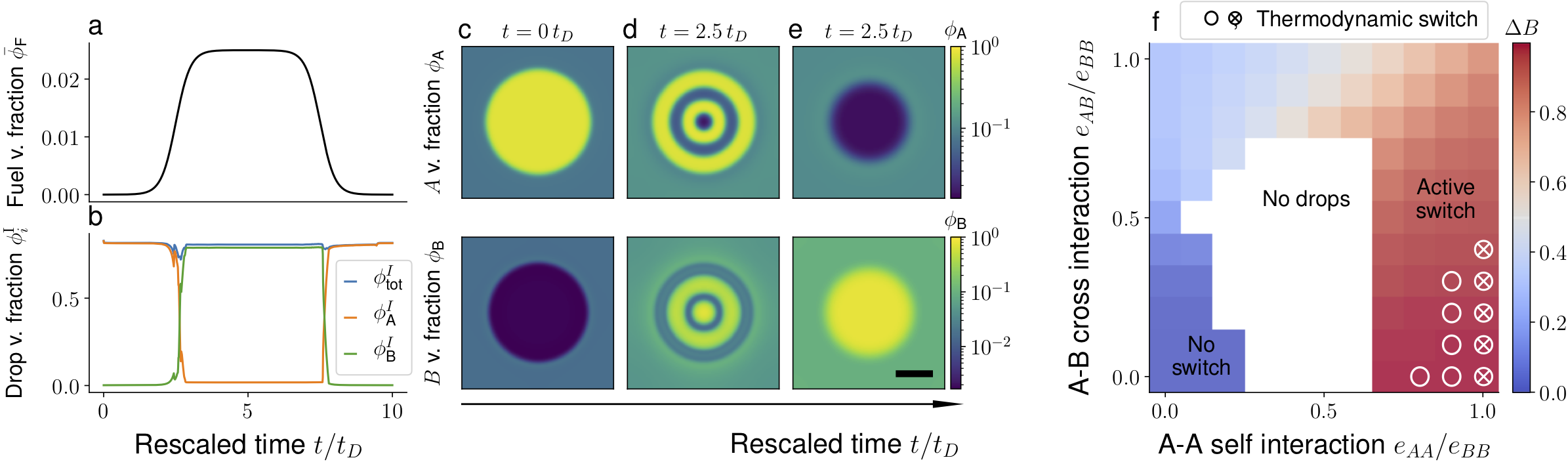}
\caption{ 
\textbf{Fuelling leads to droplet composition switch in time for a wide range of interaction parameters.}
The active switch in composition is achieved by changing fuel in time according to the protocol in (a). (b) Time traces of $\phi_{A}$, $\phi_{B}$ and $\phi_\text{tot}$ inside the dense phase show an abrupt compositional switch with time, reminiscent of the discontinuous switch observed at thermodynamic equilibrium when quasi-statically varying the temperature.
c)-e) Snapshots of $\phi_{A}$ $\phi_{B}$ profile evolving in time within the first half of the fueling ramp shown in (a). The rescaled temperature is kept constant to $T/T_0 = 0.525$.  (f) We quantify the compositional switch by determining the difference in droplet composition between (1) the stationary state (corresponding to maximum fuel value in panel (a) and (0) the initial equilibrium state: $\Delta B = \phi_B^{(1)}/\phi_\text{tot}^{(1)} - \phi_B^{(0)}/\phi_\text{tot}^{(0)}$ with different homotypic and heterotypic interactions. The active conformational switch occurs within a broader range of parameters compared to the equilibrium system referred to as ``Thermodynamic switch'' (see white symbols values).
The parameters are the same as in the lower panel of Fig.~\ref{fig:k_scan}, with fuel being constant, $T/T_0 = 0.45$ and $\phi_\text{tot} = 0.4$ and the white circles were already shown in Fig.~\ref{fig:interactions}(c). Scale bar: $20\sqrt{\kappa_A/(k_\mathrm{B}T)}$.
}
\label{fig:inver}
\end{figure}

\subsection{Active switch of droplet composition}

To test the possibility of a single droplet to kinetically swap its composition with time as the fuel is increased, we initially start with an $A$-rich droplet at thermodynamic equilibrium ($\bar{\phi}_\text{F}=0$) and gradually increase the average fuel volume fraction until it reaches a plateau value, and then gradually decrease it (Fig.~\ref{fig:inver}a). 
As the average fuel volume fraction is increased, a $B$-rich domain appears in the center of the initial $A$-rich droplet. This domain grows and splits into concentric rings enriched in $A$ and $B$, respectively (Fig.~\ref{fig:inver}d). The outermost $B$-rich domain radially propagates inwards and outwards facilitating the formation of a final $B$-rich droplet. This inversion occurs approximately concomitant to the average fuel amount exceeding some specific value  (Fig.~\ref{fig:inver}e). 
Thus, the composition has indeed swapped with time compared to the initial state. As the fuel is gradually decreased, $B$ material is consumed causing the droplet to shrink and a release of $A$ material near its interface, see SI Movie VIII. This process forms an $A$-rich outer ring that relaxes with time to the spherical shape -- the composition has swapped back to its initial value. 
These results highlight that
the composition of a droplet can indeed be controlled and reversibly switched by fuel that breaks detailed balance of the rates without dissolving and re-nucleating the droplet at another position.

The control of phase composition by fuel which we report here is in principle applicable to a rather broad class of macromolecules. In our model, their interactions are characterized by the values of their homo- and heterotypic interaction parameters. While the compositional switch controlled by temperature at thermodynamic equilibrium (``thermodynamic switch'') is only accessible within a narrow range of interaction parameters (see zoom in Fig.~\ref{fig:interactions}c), the switching via fuel (``active switch'') is possible in a significantly broader range of interaction parameters, as shown in Fig.~\ref{fig:inver}e. To characterize this broader parameter range, we calculated the jump height of the compositional change, $\Delta B = \phi_B^{(1)}/\phi_\text{tot}^{(1)} - \phi_B^{(0)}/\phi_\text{tot}^{(0)}$, for various  homo- and heterotypic interactions. These results show that for the temperature chosen, the region of the ``active switch'' is indeed much broader than the region corresponding to the``thermodynamic switch''
(Fig.~\ref{fig:inver}f).

\section{Conclusion}

In this work, we derived and analyzed a theory of a three component mixture composed of a solvent and a macromolecule that can exist in two different molecular states. Our model accounts for phase separation of such macromolecules from the solvent and reversible transitions between the two molecular states. We considered the macromolecules as polymeric molecules and thus described the interactions among all three components by a Flory-Huggins mean field free energy. 
We discussed two scenarios that essentially differ with respect to their kinetics of the molecular transition: a thermodynamic system that is fully governed by the minimization of the free energy, and a system where the flux of the molecular transition is influenced by the amount of fuel. The presence of fuel makes the transition flux independent of the model free energy and thus breaks detailed balance of the rates. In contrast to the thermodynamic system, the resulting stationary states exhibit a non-zero flux. 

Our first key finding is that, at  thermodynamic equilibrium, molecular transitions can control the composition of molecular states in both, the dilute and the dense phase as a function of temperature. Strikingly, for similarly strong self-interactions between each molecular state, we found a discontinuous switch of the dense phase between states where most macromolecules are either in one or the other state. 
Our second key finding is that a switch in composition can also be triggered in finite systems composed of macromolecule-rich droplets
using fuel -- a more likely control pathway in living cells in contrast to temperature. Most strikingly, we predict that such non-thermodynamic, fuel-related control of droplet composition is achievable within a much broader range of polymeric interaction parameters compared to the system at thermodynamic equilibrium. 

This broad parameter range renders our reported fuel-controlled, switch-like transition in droplet composition as a highly relevant mechanism to control the functionality of droplet-like condensates in living cells. Most physicochemical properties and likely all biological functionality of intra-cellular droplets are downstream consequences of the distinct local composition of the dense phase relative to its environment. Hallmark examples are the partitioning of specific molecules into protein-RNA condensates
leading for example to  increased organism fitness~\cite{franzmann2018phase}, controlled assembly hierarchies relevant for a functional tight-junction~\cite{ beutel2019phase},
and initiation of transcription and splicing machinery~\cite{guo2019pol}.
Moreover,
the droplet composition also determines the
rheological properties of condensates, e.g. its viscoelasticity~\cite{jawerth2018salt}, as well as its  slow glass-like aging kinetics~\cite{jawerth2020protein}. 
Our reported mechanism for switching droplet composition is easily accessible parameter-wise and may represent a versatile and robust kinetic pathway of controlling the droplet composition-related functionality in living cells.

From a physics perspective, our results hint toward an unexplored question, namely what determines the number of coexisting distinct domains away from thermodynamic equilibrium.
At thermodynamic equilibrium, the Gibbs phase rule sets an upper bound for the number of coexisting phases. In our specific case of an incompressible ternary mixture, the molecular transition has reduced the degrees of freedom to a single conserved variable. As a result, three-phase coexistence can only occur at a single point or line parametrized by the conserved variable. 
In contrast, in a ternary mixture without molecular transitions,
there can be an extended two-dimensional domain in the phase diagram where three phases coexist (Fig.~\ref{fig:three_p}). Thus, the molecular transition lowered the maximal dimensionality of the domain with three coexisting phases by one. 
Are there similar upper bounds in non-equilibrium systems, and if there are, can systems away from equilibrium increase or decrease this upper bound?
Given our results, breaking detailed balance of the transition rates and thus maintaining the system away from thermodynamic equilibrium releases the system from the Gibbs phase rule and allows the formation of at least three phases, and potentially even more.

\acknowledgments{
We thank J.\ Bauermann, P.\ McCall, T.\ Harmon, L.\ Hubatsch and F.\ J\"ulicher for fruitful discussions about the topic. Very special thanks goes to P.\ McCall, K.\ Alameh,  and J.\ Bauermann for very helpful feedback on the manuscript.
We thank A.\ Serrao, P.\ Schwintek , A.\ K\"uhnlein,  C.\ Mast 
and D.\ Braun for inspiring discussion on DNA gel formation and application to the Origin on Life. 
O.\ Adame-Arana acknowledges funding from the Armando and Maria Jinich postdoctoral fellowship for Mexican citizens.
G.\ Bartolucci and C.\ Weber acknowledge the SPP 2191
``Molecular Mechanisms of Functional Phase Separation'' of the German Science Foundation for financial support. 
}
 

\begin{appendix}

\section{Thermodynamic equilibrium}
\label{app:thermo_equ}

\subsection{Conserved quantities and equilibrium  condition of the molecular transition}\label{sec:red_ind_var}

To study the interplay between chemical reactions and phase separation, we consider the $T$-$P$-$N_i$-ensemble and three different species, $i=A,B,W$.
At equilibrium the total number of species $s$, the number of independent components $c$ and the number of independent chemical reactions $n_r$ are related by~\cite{Alberty2003book}
\begin{equation}
    s = c + n_r \, .
\end{equation}
In our case, we consider three species ($s=3$) and by means of one chemical reaction ($n_r=1$; see Fig.~\ref{fig:teaser}), the number of independent components gets reduced by one and thus the  dimensionality of the corresponding phase diagram reduces. As the two independent composition variables ($c=2$), we consider the quantities conserved in the reaction Fig.~\ref{fig:teaser}, namely $N_{W}$ (the number of solvent particles), and $N_\text{tot} = N_{A} + N_{B}$ (the total amount of $A$ and $B$ molecules). To find the chemical potentials corresponding to such conserved quantities, we start from the Gibbs free energy $G(T,P,N_i)$ and recall that its variation corresponding to a chemical reaction vanishes at chemical equilibrium:
\begin{equation}
     \dd G \,\big \rvert_{T,P} = \sum_i \mu_i \dd N_i = 0 \, ,
\end{equation} 
where $\mu_i(T, P, N_i) = \frac{\partial G}{\partial N_i}$. In the specific case of the molecular transition depicted in Fig.~\ref{fig:teaser}, this gives
\begin{equation}
\label{eq:chem_bal}
    \mu_{A}(T, P, N_i) = \mu_{B} (T, P, N_i) \, ,  
\end{equation}
which simplifies the variation of $G$ to 
\begin{align}
    \dd G = -S \dd T + V \dd P + \mu_{A} \dd N_\text{tot} + \mu_{W} \dd N_{W} \, .
\end{align}
This result shows that at constant temperature and pressure the Gibbs free energy only depends on the conserved variables, i.e., the total particle number $N_\text{tot}$ and the solvent particle number $N_{W}$. Moreover, we learn that the chemical potentials associated with these conserved quantities, $N_\text{tot}$ and $N_{W}$, are $\mu_{A}$ and  $ \mu_{W}$, respectively. 

To study the system at equilibrium utilizing the familiar common tangent construction (or Maxwell construction), 
we perform the Legendre transform to the
$T$-$V$-$N_i$-ensemble,  and switch to Helmholtz free energy $F = G - PV $, where the differential reads
\begin{equation}
    \dd F = -S \dd T - P \dd V + \mu_{A} \dd N_\text{tot} + \mu_{W} \dd N_{W} \, .
\end{equation}
In general, this transformation is difficult since the chemical equilibrium relation in Fig.~\ref{fig:teaser} depends on pressure and thus expressing the free energy as a function of the conserved quantity  introduces a non-trivial dependence on pressure. 

\subsection{Equilibrium condition of the molecular transitions in incompressible systems} \label{sec:cheq_incompr}

For systems where volumes vary only slightly with pressure, we can use incompressibility as a further approximation. Incompressibility is guaranteed if molecular volumes, $ \nu_i = { \frac{\partial V}{\partial N_i} }\vert_{T,P,N_{j\neq i}}$, are independent of temperature, pressure and 
composition.
While incompressibility does not reduce the number of independent variables in the $T$-$P$-$N_i$-ensemble, it provides a relation between $N_\text{tot}$, $N_{W}$, and the volume in the $T$-$V$-$N_i$-ensemble and thereby allows to further reduce the number of independent variables by one. 

However, this reduction via incompressibility requires a further condition, namely that molecular volumes are conserved in the molecular transition, Fig.~\ref{fig:teaser}. 
In our case, this corresponds to $\nu_{A} = \nu_{B}$. Thus, we obtain the following incompressibility relation connecting volume and particle numbers,  $ V = N_{A} \nu_{A} + N_{B} \nu_{B} + N_{W} \nu_{W} = N_\text{tot} \nu_{A} + N_{W} \nu_{W}$. This relationship states that variations of the total volume are caused by variations of the conserved quantities.
Using, $\nu_{A} \dd N_\text{tot}  + \nu_{W} \dd N_{W}  = \dd V $,  leads to the following differential of the Helmholtz  free energy:  
\begin{equation}
    \dd F = -S \dd T - \Pi \, \dd V + \bar{\mu}_{A}\, \dd N_\text{tot} \, .
\end{equation}
Here, we introduced the exchange chemical potential, defined as
\begin{equation}
    \label{eq:mu_ex_app}
    \bar{\mu}_i = \mu_i - \mu_{W}\frac{\nu_i}{\nu_{W}} \, ,
\end{equation} 
which characterizes the free energy difference for exchanging $A$ with solvent $W$. The pressure $\Pi=\left(P - \frac{\mu_{W}}{\nu_{W}}\right)$ is not the mechanical pressure. 
This pressure quantifies the response to variations of the total volume while $N_\text{tot}$ is kept fixed. Thus, we can think of it as the pressure acting on a semi-permeable membrane which separates the system from a solute.
This pressure  is commonly referred to as osmotic pressure.

Conservation of molecular volumes, $\nu_{A} = \nu_{B}$, allows one to rewrite the chemical equilibrium condition, Eq.~\eqref{eq:chem_bal}, in terms of the exchange chemical potentials, leading to Eq.~\eqref{eq:chem_bal_ex} in the main text.

\subsection{Free energy density and Maxwell construction} \label{sec:f}

\begin{figure}
\centering
\includegraphics[width=0.4\linewidth]{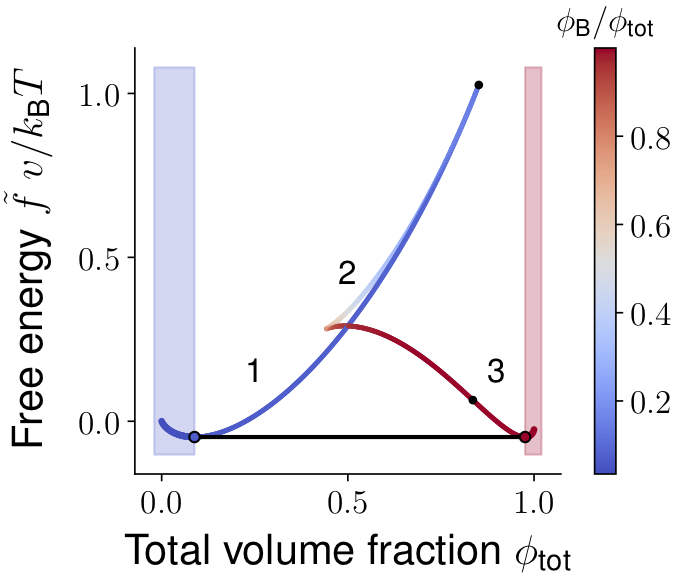}
\caption{ 
\textbf{Stable and unstable free energy branches due to molecular transitions. 
}
Free energy $\tilde f$ of the mixture as a function of the conserved quantity $\phi_\text{tot}$, where the tilde indicates that all linear terms were subtracted for illustrative purposes.
Molecular transitions can lead to the emergence of multiple solutions for a single $\phi_\text{tot}$ value in the demixing region (white). We distinguish three branches: $1$ and $3$ are metastable and $2$ is a locally unstable branch.
The common tangent (Maxwell construction) is shown in black and separates the demixing region from the mixed region (blue, red shaped). In the demixing region, the system is homogeneous, the free energy convex and all branches collapse. 
}
\label{fig:f_loop}
\end{figure}

Exploiting the chemical equilibrium condition (Eqs.~\eqref{eq:chem_curve}) and incompressibility~\eqref{eq:phitot}, we can recast the free energy in Eq.~\eqref{eq:f} in the form $f = f(T, \phi_\text{tot})$. The existence of multiple solutions of the chemical equilibrium relation in Eq.~\eqref{eq:chem_curve} can lead to the formation of a free energy loop, as displayed in Fig.~\ref{fig:f_loop}. We also show the common tangent construction (or Maxwell construction), that is used to determine the phase diagrams. For visualization purposes we display $\tilde{f} = f - m \, \phi_\text{tot}$, where $m = \left(f(\phi_\text{tot}^I)- f(\phi_\text{tot}^{II}) \right)/\left( \phi_\text{tot}^I- \phi_\text{tot}^{II}\right)$

\section{Lower transition temperature}
\label{app:trans_point}

Here we describe the criterion for the existence of the first order lower transition point. This point appears when the two solutions of the phase equilibrium coincide at $\phi_\text{tot} = 1$, i.e., when the solution becomes a polymer blend composed of $A$ and $B$ only. If the interaction entropies in Eq.~\eqref{eq:eij} are negligible, the critical temperature below which such a mixture demixes reads $k_{B}T_c = r ( e_{AA} + e_{BB} - 2\,e_{AB}) /2 $. Moreover,
incompressibility at $\phi_\text{tot} = 1$ reads
\begin{align}
    \phi^\text{I}_{A} = 1 - \phi^\text{I}_{B} = \phi^\text{II}_{B} \, ,\\
    \phi^\text{I}_{B} = 1 - \phi^\text{I}_{A} = \phi^\text{II}_{A} \, .
\end{align} 
Since the chemical potential must be equal in the two coexisting phases, for the system to phase-separate at $\phi_\text{tot}=1$, the chemical equilibrium relation must have solutions which are symmetric with respect to the exchange $\phi_{B} \rightarrow \phi_{A} = 1- \phi_{B}$, i.e.,
\begin{equation}
    \omega_{B} - \omega_{A} = e_{BB} - e_{AA} \, .
\end{equation}
With the choice $\omega_{B} - \omega_{A} = e_\text{int} - T s_\text{int}$, we find for the lower transition temperature
\begin{equation}
\label{eq:T_d}
    T_\text{d} = \frac{e_\text{int}-(e_{BB} - e_{AA})}{s_\text{int}} \, .
\end{equation}
For the phase diagram to be reentrant, this temperature must be positive. Thus, we derive the condition $e_\text{int}>e_{BB} - e_{AA}$.  Setting $e_{AA}=0$, explains the vertical line at the boundary of the ``UCST'' and ``UCST and LDT'' regions in Fig.~\ref{fig:examples}(d).
We can also derive the condition for which the upper critical point merge with the lower first order transition point, $T_\text{c} =  T_\text{d}$:
\begin{align}
    e^*_\text{int} = \left(s^*_\text{int} \,\frac{r}{2}+1\right) e^*_{BB} + \left(s^*_\text{int} \,\frac{r}{2}-1\right) e^*_{AB} - s^*_\text{int} r \,e^*_{AB} \, , 
\end{align}
which, again in the case of $e_{AA}=e_{AB}=0$, gives the line at the boundary of UCST and line in Fig.~\ref{fig:examples}(d).

\section{Linear response theory for phase separation with molecular transitions}
\label{app:non_eq}

In a system where energy is conserved, the change of the system entropy, $\dot S$, can be expressed in terms of the
integrated rate of change of the total free energy density  $f^\text{tot}= f + \sum_{i=A,B} \frac{\kappa_i}{2} |\nabla \phi_i |^2  + \frac{\kappa}{2} \nabla \phi_A \cdot \nabla \phi_B$
within the volume $V$ and
the free energy flux $J^f_\alpha$  through the volume boundaries $\partial V$~\cite{julicher2009generic}:
\begin{equation}\label{eq:total_entropy_production}
	T\dot S = - \int_V \text{d}^3x \,  \partial_t f^\text{tot} - \int_{\partial V} \text{d}\mathcal{S}_\alpha \, J^f_\alpha  \, .
\end{equation}
Using, $\partial_t f^\text{tot}=  \sum_{i=A,B} \left(\partial_t \phi_i \right)  \, \tilde{\mu}_i/\nu_i$, 
and the conservation law, $\partial_t \phi_i= -\partial_\alpha \left(j_{i,\alpha}\nu_i \right)+ s_i$, where $j_{i,\alpha}$ is the particle flux and the molecular transition between $A$ and $B$ implies ($s_i=-s_j=s$), we find for the entropy production
\begin{align}
T\dot S &= - \int_{\partial V} \text{d}\mathcal{S}_\alpha \, J^f_\alpha
 - \int_V \text{d}^3x \, s  \left( \tilde{\mu}_A/\nu_A -  \tilde{\mu}_B/\nu_B \right)
 + \int_V \text{d}^3x \, 
 \sum_{i=A,B} \tilde{\mu}_i \,  \partial_\alpha j_{i,\alpha}  \, .
 \end{align}
Partial integration leads to 
\begin{align}
T\dot S &=  - \int_V \text{d}^3x \, s  \left( \tilde{\mu}_A/\nu_A -  \tilde{\mu}_B/\nu_B \right)
 - \int_V \text{d}^3x \, 
 \sum_{i=A,B}  j_{i,\alpha}  \, \partial_\alpha \tilde{\mu}_i \, ,
 \end{align}
where we have identified the free energy flux  $J^f_\alpha= \sum_{i=A,B}  j_{i,\alpha} \, \tilde{\mu}_i$. In our special case of molecular transitions conserving the molecular volume, we have $\nu_A=\nu_B=\nu$.
Thus, if we write to linear order
\begin{subequations}
\begin{align}
s & = - \Lambda_s \,  \left( \tilde{\mu}_A -  \tilde{\mu}_B \right) \, , \\
 j_{i,\alpha} &=  - \Lambda_i \, \partial_\alpha \tilde{\mu}_i \, ,
 \end{align}
 \end{subequations}
 the system entropy will increase and 
 approach thermodynamic equilibrium. 
 Here,  ${\Lambda}_s $ denotes the mobility for the molecular transition
 and $\Lambda_i$ is the diffusive mobility, and we neglect cross couplings for simplicity. Moreover,
 we have used $\nu_A=\nu_B=\nu$ and absorbed it into the definition of the mobility ${\Lambda}_s$. 
The equilibrium conditions are $\tilde{\mu}_i=\text{const.}$ and $\tilde{\mu}_A=\tilde{\mu}_B$. Note that for an infinitely large thermodynamic system, 
we can neglect the spatial derivative in the chemical potentials, and express the equilibrium conditions in terms of the exchange chemical potentials  
$\bar{\mu}_i=\text{const.}$
and $\bar{\mu}_A=\bar{\mu}_B$ (Eq.~\eqref{eq:chem_bal_ex}).

\section{Chemical flux for molecular transitions with broken detailed balance}
\label{app:brok_det_bal}

When detailed balance of the molecular transition rates is broken and considering a non-linear dependence of the chemical flux on the chemical potentials, the chemical flux in Eq.~\eqref{eq:s_fuel} can be written as~\cite{Weber2019,Zwicker2017}
\begin{align}
s &= \Lambda_s \left[ \exp \left( \frac{ (\tilde{\mu}_{A}-\tilde{\mu}_{B}) + \Delta \mu}{k_{B}T}  \right)-1\right] \, .
\end{align}
Here, we imposed a chemical potential $\Delta \mu \not=0$ that makes sure that the system cannot reach thermodynamic equilibrium with $\tilde{\mu}_{A} =\tilde{\mu}_{B}$ (Eq.~\eqref{eq:chem_bal_ex}). For homogeneous systems which obey $\bar{\mu}_{A}-\bar{\mu}_{B} + \Delta \mu < k_{B}T$, the leading order of the chemical flux can be written as 
\begin{align}
s& \simeq \Lambda_s    \frac{ (\tilde{\mu}_{A}-\tilde{\mu}_{B}) }{k_{B}T}  + \Lambda_s \frac{\Delta \mu}{k_{B}T} \, .
\end{align}
In the following, we impose the 
 chemical potential by a fuel component of volume fraction $\phi_\text{F}$, and expand $\Delta \mu(\phi_{A},\phi_{B},\phi_\text{F})$ to lowest order ($\phi_i \ll \phi_W$):
 \begin{equation}
 	\Delta \mu \simeq (K_\leftarrow \phi_{B} - K_\rightarrow \phi_{A}) \phi_\text{F} \, .
\end{equation}	
Defining 
$k_{\rightarrow} =  \Lambda_s \frac{K_\rightarrow}{k_{B}T}$
and
 $k_\leftarrow =  \Lambda_s \frac{K_\leftarrow}{k_{B}T}$, we get the chemical flux shown in  Eq.~\eqref{eq:s_fuel}.

\section{Derivation of equilibrium fuel profile}
\label{app:fuel_eq_part}

We impose fuel conservation by keeping the spatial average of Eq.~\eqref{eq:phi_fuel} fixed and equal to $\bar{\phi}_\text{F}$. This implies $\alpha + \beta \bar{\phi}_\text{tot} = 1$, with $\bar{\phi}_\text{tot}$ being the total average volume fraction of macromolecules (see Eq.~\eqref{eq:phitot}). In Eq.~\eqref{eq:phi_fuel} we notice the coefficient $\beta$ encodes correlations between fuel $\phi_\text{F}$ and total macromolecular material $\phi_\text{tot}$. Maximal spatial correlation between $\phi_\text{F}$ and $\phi_\text{tot}$ is reached maximizing $\beta$ with the constraints $\alpha + \beta \bar{\phi}_\text{tot} = 1$ and $0<\phi_\text{F}(x)<1$ everywhere in space. This leads to $\alpha =0$ and $ \beta=1/\bar{\phi}_\text{tot}$.
Maximal anti-correlation between $\phi_\text{F}$ and $\phi_\text{tot}$ is reached minimizing $\beta$ with the same constraints, leading to $\alpha = -\beta = 1/(1-\bar{\phi}_\text{tot})$. Finally, no correlation between $\phi_\text{F}$ and $\phi_\text{tot}$, i.e. fuel homogeneously distributed in the system, is achieved for $\beta=0$ and, due to fuel conservation, $\alpha=1$. This explains the choices of $\alpha$ and $\beta$ introduced at the end of Sec.~\ref{sec:non_eq_thermo}.

At equilibrium and for the case where the fuel has only weak effects on the chemical flux (i.e., $k_\leftarrow\simeq 0, k_\rightarrow \simeq 0$), we can make the connection between the coefficients $\alpha$ and $\beta$ in Eq.~\eqref{eq:phi_fuel} and the fuel partitioning even more explicit. We recall the definition of partitioning coefficient of the fuel component, $P_\text{F}=\phi_\text{F}^\text{I}/\phi_\text{F}^\text{II}$, and of the total concentration, $P_\text{tot}=\phi_\text{tot}^\text{I}/\phi_\text{tot}^\text{II}$.
Here,
I and II denote the dense and the  dilute phase, respectively. 
We can express the fuel and total volume fractions in I and II, respectively,
with the average fuel volume fraction $\bar{\phi}_\text{F}$
 that is considered to be maintained at some constant value, and the conserved total volume fraction $\bar{\phi}_\text{tot}$:
\begin{align}
\phi_i^\text{I} &= \xi_i \,  P_i \,  \bar{\phi}_\alpha \, , \\
\phi_i^\text{II}& = \xi_i \bar{\phi}_\alpha\, ,
\end{align}
where $i=\{ \text{F},\text{tot} \}$. The partition degree~\cite{weber2019spatial} reads
\begin{equation}
\xi_i = \frac{1}{1+(P_i -1) \frac{V^\text{I}}{V}}\, ,
\end{equation}
where the phase-separated volume in the limit of dilute fuel is given as
\begin{equation}
{V^\text{I}} = V \frac{\bar{\phi}_\text{tot} - \phi_\text{tot}^\text{II} }{\phi_\text{tot}^\text{I} - \phi_\text{tot}^\text{II} } \, .
\end{equation}
Evaluating Eq.~\eqref{eq:phi_fuel} inside and outside the dense phase, we find: 
\begin{align}
&\alpha = 
\xi_\text{F}  \left( 1- 
\frac{ P_\text{F}-1  }{P_\text{tot}-1 }\right) \, ,\\
&\beta =  \frac{\xi_\text{F} }{\xi_\text{tot} }  \frac{1}{\bar{\phi}_\text{tot}}
\frac{ P_\text{F}-1  }{P_\text{tot}-1 } \, .
\end{align}
If the fuel partitions equally strong into both phases ($P_\text{F}=1$, and thus  $\xi_\text{F}=1$), we get $\alpha= 1$ and $\beta=0$. Consistently, this corresponds to a homogeneous fuel profile, $\phi_\text{F}(x) = \bar{\phi}_\text{F}$. For a fixed $P_\text{tot}>1$, the fuel partition coefficient $P_\text{F}$ determines the localization of the fuel. In particular, $P_\text{F} >1$, corresponds to fuel co-localizing with the total volume fraction $\phi_\text{tot}$ with $\beta>0$. On the contrary, when $ P_\text{F} <1$,  the fuel and the total volume fraction $\phi_\text{tot}$ anti-co-localize with  $\beta<0$.

\section{Numerical methods, interface tracking and parameter choices}
\label{app:kin_and_numerics}

We solve the kinetic equations~\eqref{eq:phiAphiB} and~\eqref{eq:s_fuel} in two dimensions with periodic boundary conditions. First, we use the energy quadratization method~\cite{YANG2017104, Review_EQ_2018, ZHAO2019382} to map the free energy of the system into a quadratic form. Then, we use the second-order finite difference method in space and the Crank-Nicolson method in time to discretize the partial differential equations. A stabilizing term~\cite{Shen2010} is added allowing larger time steps.

The interfaces between $A$-rich, $B$-rich, and solvent-rich domains, shown by dashed red and blue lines in Fig.~\ref{fig:k_scan} (a) and (f), are defined as the contour lines of the functions $\phi_A(x)$ and $\phi_B(x)$ respectively,  corresponding to the value $\phi_\text{tot}^*=(\phi_\text{tot}^\text{max}-\phi_\text{tot}^\text{min})/2$. Here,  $\phi_\text{tot}^\text{max}$ and $\phi_\text{tot}^\text{min}$ correspond to the maxinum and the minimum value of $\phi_\text{tot}(x)$, for a given time. When, instead, we are interested in identifying $\phi_\text{tot}$-rich and $\phi_\text{tot}$-poor phases, like in Fig.~\ref{fig:inver} (b), without distinguishing between $A$ and $B$ conformations, we simply compare $\phi_\text{tot}(x)$ with $\phi_\text{tot}^*$.

We measure the effective rates in units of $t_D= L^2 / D_A $,  which represents the time an $A$ particle takes  to diffuse across a length $L = 100 \sqrt{\kappa_A/(k_\mathrm{B}T) }$, where $L$ corresponds to the system size. Moreover, the length scale $\sqrt{\kappa_A/(k_\mathrm{B}T) }$ approximates the droplet interface width~\cite{Weber2019}. In all simulations, we set $\Lambda_s = 5 \, t_D^{-1} $.

For the first prototypical parameter set (i) ($e_{AA} = 0.2 \, e_{BB}$, $ e_{AB} = 0 $), we have chosen 
$\kappa_A = \kappa_B$ and $ \kappa = 0$. For the second  prototypical parameter set (ii) ($e_{AA} = e_{BB}$, $ e_{AB} = 0.2 $), we have chosen $\kappa_B = 5 \, \kappa_A $ and $ \kappa = 2 \, \kappa_A$. This choice is motivated by the possibility of having active droplet inversion without dissolving the droplet and subsequently re-nucleating it. 
In particular, the hierarchy $\kappa_B > \kappa > \kappa_A$ favors wetting of an $A$-rich layer onto a $B$-rich droplet. This is important in the second half of the fuel ramp (see Fig.~\ref{fig:inver} (a)), in order to produce $A$ material not too far from the shrinking $B$-rich droplet, see SI movie VIII.

We obtained solutions to Eqs.~\eqref{eq:phiAphiB} for a total time period up to $40 \, t_D$. Within this time, the majority of patterns shown in Fig.~\ref{fig:k_scan} appeared to be stable and stationary.   

\end{appendix}

\medskip

\bibliography{ms}

\end{document}


\title{
Supplementary Information: Controlling composition of coexisting phases via
molecular transitions
}

\author{Giacomo Bartolucci}
\affiliation{Max Planck Institute for the Physics of Complex Systems,
N\"{o}thnitzer Strasse~38, 01187 Dresden, Germany}
\affiliation{Center for Systems Biology Dresden,  Pfotenhauerstrasse~108, 01307 Dresden, Germany}

\author{Omar Adame-Arana}
\affiliation{Department of Chemical and Biological Physics, Weizmann Institute of Science, Rehovot 76100, Israel}
\author{Xueping Zhao}
\affiliation{Max Planck Institute for the Physics of Complex Systems,
N\"{o}thnitzer Strasse~38, 01187 Dresden, Germany}
\affiliation{Center for Systems Biology Dresden,  Pfotenhauerstrasse~108, 01307 Dresden, Germany}

\author{Christoph A.\ Weber\footnote{\label{CA}Corresponding author.}}
\affiliation{Max Planck Institute for the Physics of Complex Systems,
N\"{o}thnitzer Strasse~38, 01187 Dresden,
Germany}
\affiliation{Center for Systems Biology Dresden,  Pfotenhauerstrasse~108, 01307 Dresden, Germany}

\date{\today}


\maketitle

\section*{Movie description}

\begin{figure*}
\centering
 \includegraphics[width=\linewidth]{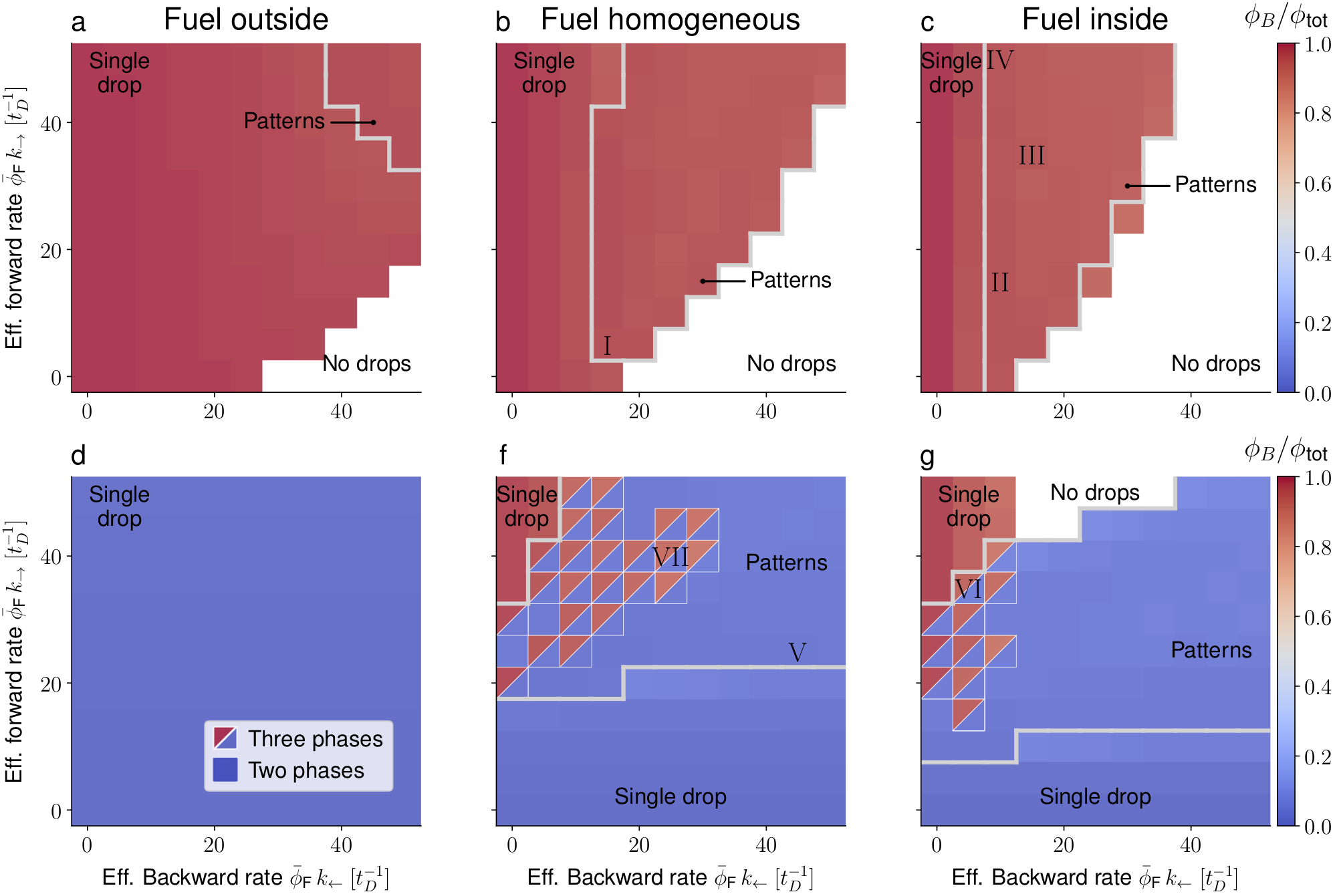}
\caption{
\textbf{Molecular transition breaking detailed balance leads to non-equilibrium stationary states.}
We study the emergence of non-equilibrium stationary states due to the presence of the fuel as a function of the effective forward and backward rates. We tackle both weak self interactions of the $A$ species ,$e_{AA} = 0.2 \, e_{BB}$, $ e_{AB} = 0 $, (a-c), and strong $A$-$A$ self interaction ,$e_{AA} = e_{BB}$, $ e_{AB} = 0.2e_{BB} $, (d-g). We display the values of the effective rates corresponding to the selected movies I-VIII. 
Here the color code indicates the dense clusters composition $\phi_{B}/\phi_\text{tot}$. For strong $A$-$A$ self interactions, stationary patterns composed of three distinct domains appear. In the diagrams (d)-(g) they are indicated with with two triangles of different colors, that correspond to the composition of the two distinct dense phases. For details see the main text.}
\label{fig:SI_k_scan}
\end{figure*}

Parameter set (i), weak self interactions of the $A$ species ($e_{AA} = 0.2 \, e_{BB}$, $ e_{AB} = 0 $):
\newline
\textbf{Movie I:} The initial equilibrium drop changes to a ring. For this movie the fuel is homogeneous in space and the effective rates are $\bar{\phi}_\text{F} \, k_\leftarrow =15  \,t_D^{-1}$ and $\bar{\phi}_\text{F} \, k_\rightarrow =5 \,t_D^{-1}$.
\newline
\textbf{Movie II:} The initial equilibrium drop evolves into smaller drops surrounded by an outer ring. For this movie the fuel partitions inside the dense phase and the effective rates are $\bar{\phi}_\text{F} \, k_\leftarrow =10 \,t_D^{-1}$ and $\bar{\phi}_\text{F} \, k_\rightarrow =15 \,t_D^{-1}$.
\newline
\textbf{Movie III:} The initial equilibrium drop splits up into a ring and multiple drops. The ring get unstable wit time and a pattern composed of equally sized drops evolves.  For this movie the fuel partitions inside the dense phase and the effective rates are $\bar{\phi}_\text{F} \, k_\leftarrow =15 \,t_D^{-1}$ and $\bar{\phi}_\text{F} \, k_\rightarrow =35 \,t_D^{-1}$.
\newline
\textbf{Movie IV:} The initial equilibrium drop breaks into smaller fragments, subsequently three of them elongate and align, a bit reminiscent of stripes. For this movie the fuel partitions inside the dense phase and the effective rates are $\bar{\phi}_\text{F} \, k_\leftarrow =10 \,t_D^{-1}$ and $\bar{\phi}_\text{F} \, k_\rightarrow =50 t_D^{-1}$.

\vspace{5mm}

Parameter set (ii), strong self interactions of the $A$ species ($e_{AA} = e_{BB}$, $ e_{AB} = 0.2e_{BB} $):
\newline
\textbf{Movie V:} When the fuel is injected to the system holes appear in the drop. For this movie the fuel is constant in space $\bar{\phi}_\text{F} \, k_\leftarrow =45 \,t_D^{-1}$ and $\bar{\phi}_\text{F} \, k_\rightarrow =25  \,t_D^{-1}$.
\newline
\textbf{Movie VI:} When the fuel is injected to the system composed of an initial $A$-rich drop, an internal $A$-rich ring forms and then becomes a central drop. The remaining $A$ material forms drops that initially fill the space but, for later times, droplets far from the central $B$-rich drops dissolve. For this movie the fuel partitions inside the droplet and the effective rates are $\bar{\phi}_\text{F} \, k_\leftarrow =5 \,t_D^{-1}$ and $\bar{\phi}_\text{F} \, k_\rightarrow =35 \,t_D^{-1}$.
\newline
\textbf{Movie VII:} When the fuel is injected to the system, in the center of the initial $A$-rich drop an internal $B$-rich drop forms. The remaining $A$ material organizes into a ring wetting the $B$-rich central drop, and an outer layer of multiple drops. For this movie the fuel partitions inside the droplet and the effective rates are $\bar{\phi}_\text{F} \, k_\leftarrow =25 \,t_D^{-1}$ and $\bar{\phi}_\text{F} \, k_\rightarrow =40 \,t_D^{-1}$.
\newline
\textbf{Movie VIII:} When the fuel is injected to the system, in the center of the initial $A$-rich drop becomes $B$-rich. When subsequently fuel is taken out the droplet switches back to its original composition. For this movie the is homogeneous in space and the effective rates are $\bar{\phi}_\text{F} \, k_\leftarrow =2.5 \, 10^{-5} \,t_D^{-1}$ and $\bar{\phi}_\text{F} \, k_\rightarrow =4 \, 10^{-4} \,t_D^{-1}$.